\documentclass[sigconf, authorversion]{acmart}
\usepackage{acronym}
\usepackage[]{todonotes}
\usepackage{xspace}
\usepackage{multirow}
\usepackage{amsmath}
\usepackage{hyperref}
\usepackage{comment}
\usepackage{listings}
\usepackage{tabularx}
\usepackage{booktabs}
\usepackage{makecell}
\usepackage{pifont}
\usepackage{url}
\usepackage{breakurl}
\usepackage{color, colortbl}

\usepackage{float}
\usepackage{tikz}
\usetikzlibrary{positioning, fit, calc, shapes, arrows}
\usepackage[underline=false]{pgf-umlsd}
\usepackage{tikz-uml}
\usepackage{listings}
\usepackage{color}
\usepackage[subtle]{savetrees}
\definecolor{listinggray}{gray}{0.92}

\AtBeginDocument{%
  \providecommand\BibTeX{{%
    \normalfont B\kern-0.5em{\scshape i\kern-0.25em b}\kern-0.8em\TeX}}}

\copyrightyear{2021}
\acmYear{2021}
\setcopyright{rightsretained}
\acmConference[WiSec '21]{14th ACM Conference on Security and Privacy
in Wireless and Mobile Networks}{June 28-July 2, 2021}{Abu Dhabi, United
Arab Emirates}
\acmBooktitle{14th ACM Conference on Security and Privacy in Wireless
and Mobile Networks (WiSec '21), June 28-July 2, 2021, Abu Dhabi, United
Arab Emirates}\acmDOI{10.1145/3448300.3467821}
\acmISBN{978-1-4503-8349-3/21/06}

\acrodef{GPS}{Global Positioning System}
\acrodef{DoS}{Denial-of-Service}
\acrodef{PoC}{Proof-of-Concept}
\acrodef{TD}{Tracking Device}
\acrodef{CD}{Communication Device}
\acrodef{TS}{Tracking Service}
\acrodef{TApp}{Tracking App}
\acrodef{BLE}{Bluetooth Low Energy}
\acrodef{TLS}{Transport Layer Security}
\acrodef{NSC}{\emph{Necessary security condition}}
\acrodef{TEE}{Trusted Execution Environment}

\newcommand{\etc}{\textit{etc.}}
\newcommand{\eg}{\textit{e.g.}}
\newcommand{\ie}{\textit{i.e.}}

\newcommand{\tbl}[1]{\ensuremath{\sf Table~\ref{#1}}}
\newcommand{\fig}[1]{\ensuremath{\sf Figure~\ref{#1}}}
\newcommand{\sect}[1]{\ensuremath{\sf Section~\ref{#1}}}

\newcommand{\capone}{\ensuremath{C_{1}}}
\newcommand{\captwo}{\ensuremath{C_{2}}}
\newcommand{\capthr}{\ensuremath{C_{3}}}
\newcommand{\capfour}{\ensuremath{C_{4}}}
\newcommand{\capfive}{\ensuremath{C_{5}}}
\newcommand{\capsix}{\ensuremath{C_{6}}}
\newcommand{\capseven}{\ensuremath{C_{7}}}
\newcommand{\capeight}{\ensuremath{C_{8}}}

\newcommand{\systemname}{\textsc{SECrow}\xspace}

\newcommand{\tabitem}{~~\llap{\textbullet}~~}
\newcolumntype{L}[1]{>{\raggedright\let\newline\\\arraybackslash\hspace{0pt}}m{#1}}
\newcolumntype{R}[1]{>{\raggedleft\let\newline\\\arraybackslash\hspace{0pt}}m{#1}}
\newcolumntype{C}[1]{>{\centering\let\newline\\\arraybackslash\hspace{0pt}}m{#1}}
\newcolumntype{Z}[0]{>{\centering\let\newline\\\arraybackslash\hspace{0pt}}X}%

\newcounter{an}
\newcounter{am}
\newcounter{cm}

\newcommand{\mypar}[1]{\vspace{1.0pt}\noindent\textbf{#1}\xspace}

\newcommand{\todoan}[2][]{%
	\refstepcounter{an}%
	{%
		\todo[color={green!30},inline]{%
			\textbf{Andrea [\uppercase{#1}\thean]:}~#2}%
	}}

\newcommand{\bloodymess}[7][0]{
  \stepcounter{seqlevel}
  \path
  (#2)+(0,-\theseqlevel*\unitfactor-0.7*\unitfactor) node (mess from) {};
  \addtocounter{seqlevel}{#1}
  \path
  (#4)+(0,-\theseqlevel*\unitfactor-0.7*\unitfactor) node (mess to) {};
  \draw[->,>=angle 60] (mess from) -- (mess to) node[midway, above]
  {\ensuremath #3};

  \if R#5
    \node (#3 from) at (mess from) {\llap{{\ensuremath #6}~}};
    \node (#3 to) at (mess to) {\rlap{~{\ensuremath #7}}};
  \else\if L#5
         \node (#3 from) at (mess from) {\rlap{~{\ensuremath #6}}};
         \node (#3 to) at (mess to) {\llap{{\ensuremath #7}~}};
       \else
         \node (#3 from) at (mess from) {\ensuremath #6};
         \node (#3 to) at (mess to) {\ensuremath #7};
       \fi
  \fi
}

\newcommand\encircle[1]{%
  \tikz[baseline=(X.base)] 
    \node (X) [draw, shape=circle, inner sep=0] {\strut #1};}

\begin{document}


\title{Toward a Secure Crowdsourced Location Tracking System}

%
%
%
%
%
%
\author{Chinmay Garg}
\affiliation{\institution{University of California, Santa Barbara}}
\email{chinmay@ucsb.edu}

\author{Aravind Machiry}
\affiliation{\institution{Purdue University}}
\email{amachiry@purdue.edu}

\author{Andrea Continella}
\affiliation{\institution{University of Twente}}
\email{acontinella@iseclab.org}

\author{Christopher Kruegel}
\affiliation{\institution{University of California, Santa Barbara}}
\email{chris@cs.ucsb.edu}

\author{Giovanni Vigna}
\affiliation{\institution{University of California, Santa Barbara}}
\email{vigna@cs.ucsb.edu}
%


\begin{abstract}
	Low-energy Bluetooth devices have become ubiquitous and widely used for different applications.
Among these, Bluetooth trackers are becoming popular as they allow users to track the location of their physical objects.
To do so, Bluetooth trackers are often built-in within other commercial products connected to a larger crowdsourced tracking system.
Such a system, however, can pose a threat to the security and privacy of the users, for instance, by revealing the location of a user's valuable object.
In this paper, we introduce a set of security properties and investigate the state of commercial crowdsourced tracking systems, which present common design flaws that make them insecure.
Leveraging the results of our investigation, we propose a new design for a secure crowdsourced tracking system (\systemname{}), which allows devices to leverage the benefits of the crowdsourced model without sacrificing security and privacy.
Our preliminary evaluation shows that~\systemname is a practical, secure, and effective crowdsourced tracking solution.

\end{abstract}

\maketitle

\section{Introduction}

Tracking devices have become very common for a wide range of applications and use cases.
People use tracking devices to monitor trucking fleet movements, commercial and defense vehicles, monitor their pets and keep track of the location of their valuables in case they are stolen or lost.
While trackers can rely on different technologies, such as radio waves, GPS, and WiFi, Bluetooth Low-Energy (BLE) trackers have become the main choice for consumer-targeted trackers~\cite{chipolotr,tile,trackr}, and they are currently used by millions of users.
For instance, the popular TrackR app has more than 1,000,000 downloads from the Google Play Store~\cite{trackrapp}.

In practice, BLE trackers consist of small and cheap devices that can be attached to key rings, luggage, wallets, or any other personal belonging that a user wishes to track.
Once activated, they work by communicating with a mobile app installed on the users' smartphone. The app allows the users to interact with the trackers (e.g., make it ring) and acts as a medium of communication to ``connect'' the BLE trackers to the Internet.
In fact, the ``killer feature'' of such devices is the ability for users to remotely query for the location of their valuables.
To do so, BLE trackers are embedded into a more complex ecosystem, called \emph{crowdsourced location tracking system}.
In a crowdsourced model, users can track the location of their physical devices by leveraging information collected and produced by \emph{other} users in the network.
In practice, crowdsourced tracking systems rely on all their users, and on their smartphones' GPS, to obtain information about the location of their nearby BLE trackers.
This allows the owner of a BLE tracker to know the location of their valuable (almost) in real-time, even when they are not in proximity of the tracker.
At the same time, the crowdsourced model allows vendors to drastically reduce the cost of their trackers, as they do not need to embed GPS capabilities.

However, due to the nature of the distributed crowdsourced model, which utilizes smartphones for location updates, crowdsourced location tracking systems raise questions regarding the security and privacy of their protocols.
Such questions come naturally since attackers could potentially abuse the systems to perform illicit operations, \eg, stealing or inferring the location of a valuable object.
Indeed, a recent study showed that different tracking systems led to various privacy issues~\cite{weller2020lost}.

In this work, we perform a detailed, principled, and comprehensive security analysis of the existing crowdsourced location tracking systems.
%
We first formalize a set of security properties that every crowdsourced tracking system should fulfill, together with the conditions that must be satisfied to guarantee such properties.
We then study the most popular, commercial, crowdsourced tracking systems, identifying major systematic flaws that violate our security properties and, therefore, allow attackers to obtain unauthorized access to tracked devices.
We responsibly disclosed our findings to the affected vendors.

Finally, inspired by our security properties, we design the protocol of a secure crowdsourced tracking system (\systemname{}), which allows users to leverage all the benefits of the crowdsourced model without compromising their security and privacy.
Our design guarantees reliable (i.e., unspoofable) and anonymous location information, and provides end-to-end encryption, where only the owner of a BLE tracker can access its location. 
As a consequence, our design protects the location information even from server-side data breaches~\cite{continella18:bucketsec,capitalone_breach,zuo2019does}.

In summary, we make the following contributions:

\begin{itemize}

    \item We investigate the security issues arising from the adoption of the crowdsourced model for location tracking.
    We define and formalize a set of security properties and conditions that crowdsourced location tracking systems must guarantee.

    \item We perform a security analysis of the most popular, commercial, crowdsourced tracking systems, demonstrating a real threat for users.
    Specifically, by analyzing and instrumenting their mobile apps, we identify major design flaws in all these systems.
    We reported our findings to the responsible entities. 

    \item We design \systemname, a secure protocol for crowdsourced location tracking that fulfills all our security properties, protecting users from various attacks, and providing end-to-end encryption.
    We show the practicality of our approach by implementing and testing a prototype.

\end{itemize}

In the spirit of open science, we make all our code available at \url{https://github.com/ucsb-seclab/SECrow}.

\section{Background}
\begin{figure}[t]
    \centering
    \includegraphics[width=0.95\columnwidth]{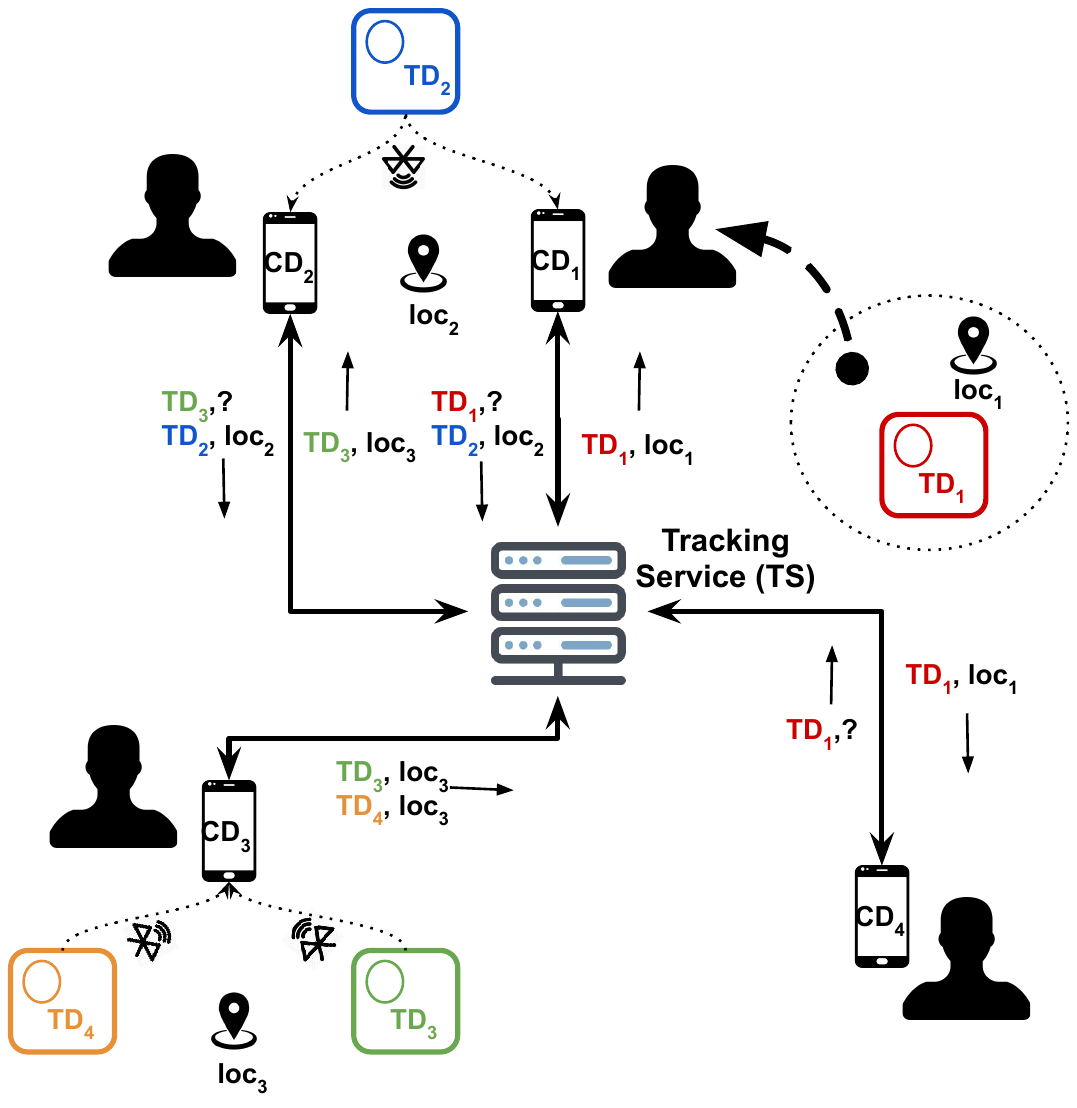}
    \vspace{-0.4cm}
    \caption{Data-flow of a representative crowdsourced tracking system.
    $\ac{CD}_{1}$ and $\ac{CD}_{4}$ own $\ac{TD}_{1}$.
    $\ac{CD}_{2}$ owns $\ac{TD}_{2}$ and $\ac{TD}_{3}$.
    $\ac{CD}_{3}$ owns $\ac{TD}_{4}$.
    The dashed arrow indicates that the user with $\ac{CD}_{1}$ moved from location $loc_{1}$ to $loc_{2}$.
    }
    \vspace{-0.3cm}
    \label{fig:crowdsourced-tracking-architecture}
\end{figure}

In this section, we explain crowdsourced tracking using common terms that will be used throughout the paper.
The goal of crowdsourced tracking is to provide users with the ability to track the location of physical devices by leveraging the crowdsourcing model, \ie, by leveraging information collected and produced by other users in the network.
A typical crowdsourced tracking system involves three entities (Figure~\ref{fig:crowdsourced-tracking-architecture}).

\begin{itemize}

\item \textbf{\acp{TD}:} The actual devices being tracked (\eg{}, TrackR~\cite{trackr}, CUBE~\cite{cubetr}).
These devices are battery-powered, communicate using~\ac{BLE}, and they are compact so that they can be attached to other physical objects, such as wallets and keys.
This provides the ability to track the corresponding physical objects by leveraging the attached~\ac{TD}.
\item \textbf{\ac{TS}:} A web service (usually REST-based~\cite{masse2011rest}) provided by the manufacturer of~\acp{TD} that allows users to know the location of their~\acp{TD}.

\item \textbf{\ac{CD}:} A device that can directly communicate with a~\ac{TD}.
Most commonly, this is an Internet-enabled smartphone running an app, which we call~\ac{TApp}, provided by the~\ac{TD} vendors.
The~\ac{TApp} can communicate with the~\ac{TS} and~\acp{TD}.
If the~\ac{CD} has location service capabilities, then the~\ac{TApp} can update~\ac{TS} with the current location of all the~\ac{BLE}-visible~\acp{TD}, \ie{}, those that are nearby.
This information from~\acp{CD} represents the crowd input.
The information from all the~\acp{CD} is used by the~\ac{TS} to provide the ability for users to know the location of their~\acp{TD}.
The~\ac{TApp} also provides an easy way for users to track~\acp{TD} and thus provides an implicit incentive for the users to install the app, thereby participating in crowdsourced tracking.
\end{itemize}

\mypar{Owners.} Every~\ac{TD} is owned by one or more users who can perform privileged operations on it, such as make it ring and flash.
The ownership is established by registering and claiming the device through a pairing process, where the~\ac{TApp} adds the ~\ac{TD} as a registered item within the user account.
Generally, users use~\ac{TApp} to interact with owned~\acp{TD} through~\ac{BLE}.
Furthermore, owners can query the last known location of a~\ac{TD}, that is~\emph{lost or not in~\ac{BLE} proximity}, by communicating with the~\ac{TS}---through the~\ac{TApp}.

A representative crowdsourced tracking system is depicted in \fig{fig:crowdsourced-tracking-architecture}, which shows four~\acp{TD}, four~\acp{CD}, and the~\ac{TS}:
$\ac{CD}_{1}$ and $\ac{CD}_{4}$ own $\ac{TD}_{1}$;
$\ac{CD}_{2}$ owns $\ac{TD}_{2}$ and $\ac{TD}_{3}$;
$\ac{CD}_{3}$ owns $\ac{TD}_{4}$.
The dashed arrow indicates that the user with $\ac{CD}_{1}$ moved from location $loc_{1}$ to $loc_{2}$.

\mypar{Location update.} All~\acp{CD} constantly update the~\ac{TS} with the location of the~\acp{TD} that are in~\ac{BLE} proximity.
Specifically, $\ac{CD}_{1}$ and $\ac{CD}_{2}$ update the location of $\ac{TD}_{2}$ using the tuple $\ac{TD}_{2}$, $loc_{2}$. Similarly, $\ac{CD}_{3}$ updates the location of $\ac{TD}_{3}$ and $\ac{TD}_{4}$.
$\ac{CD}_{4}$ does not send any location update since there are no~\acp{TD} nearby.

\mypar{Location query.} The owners of a~\ac{TD} can query~\ac{TS} about the location of their~\ac{TD}.
The~\ac{TS} responds with the last known location of the~\ac{TD}.
In~\fig{fig:crowdsourced-tracking-architecture}, we can see that $\ac{CD}_{1}$, $\ac{CD}_{2}$, and $\ac{CD}_{4}$ issue location queries for $\ac{TD}_{1}$, $\ac{TD}_{3}$, and $\ac{TD}_{1}$ respectively represented by a tuple of the form $\ac{TD}_{x}, ?$.
\ac{TS} responds to such location queries with the last known location of the \acp{TD} in the form $\ac{TD}_{x}$, $loc_{y}$.
\acp{CD} do not need to send location queries for their~\acp{TD} when they are nearby, as they can communicate directly through \ac{BLE}.
For instance, $\ac{CD}_{3}$ does not send any location query for $\ac{TD}_{4}$ as it is~\ac{BLE} visible to the corresponding $\ac{CD}_{3}$.

\section{Security properties}
\label{subsub:securityprop}
As already mentioned, the main goal of any crowdsourced tracking system is to provide the ability for users to reliably obtain the location of their tracked devices.
From a security and privacy perspective, we also want such systems to guarantee certain properties, \eg, only owners can perform privileged operations. 
Furthermore, for each property, we identify the~\emph{necessary conditions} that need to be satisfied to guarantee the corresponding property.
We call these the~\acp{NSC} of a security property.
This can be represented formally as,
\begin{equation*}
\forall C \in NSCs(S) : \lnot C \rightarrow \lnot S.
\label{lab:secp}
\end{equation*}
where $NSCs(S)$ is the set of all the necessary conditions for the security property $S$.
This formulation implies that if any of the~\acp{NSC} does not hold, then the corresponding security property is violated, providing an effective method to check for \emph{violations} of the security properties.
All the security properties are named as $X-S_{y}$, where $X$ is the entity the security property belongs to, and $y$ is a number indicating a unique id.
Similarly, all the necessary conditions are named as $C_{n}$, where $n$ is a positive number.
Next, we list all the security properties and their corresponding necessary conditions.

\vspace{-0.1cm}
\subsection{\ac{TD} security properties}
\label{subsub:tdproperties}

\mypar{Privileged owners ($TD-S_{1}$).} Only owners can perform privileged operations on a \ac{TD}. Specifically, we consider the following operations as privileged: 

	\begin{itemize}
	\item Obtaining the location of the~\ac{TD} from the~\ac{TS}.
	\item Making the \ac{TD} ring or blink remotely (\ie{}, through~\ac{BLE}) to capture human attention.
	\end{itemize}

To enforce the above restrictions, it is important for a~\ac{TD} and the~\ac{TS} to have an ability to~\emph{recognize the owners} of the~\ac{TD} to restrict access to the privileged operations.
Hence, we see that following as the~\acp{NSC} for the $TD-S_{1}$ security property:

\begin{itemize}
    \item \textbf{\capone{}:} The~\ac{TD} must able to recognize its owners.
    \item \textbf{\captwo{}:} The~\ac{TS} must able to recognize the owners of a given~\ac{TD}.
\end{itemize}

\mypar{Physical ownership ($TD-S_{2}$).} It must be impossible to own a tracking device without physical access.
This is based on the traditional ownership requirement~\ie{} to own a device, it is important to verify that the device can be accessed physically.
To ensure this, we need the ability to verify physical ownership.
Consequently, we need the following~\acp{NSC} for the $TD-S_{2}$ security property:

\begin{itemize}
    \item \textbf{\capthr{}:} Owning of a~\ac{TD} must involve performing a physical action on the~\ac{TD}. 
    \item \textbf{\capfour{}:} Registering as a owner of a~\ac{TD} with the~\ac{TS} must involve communicating with the~\ac{TD}.
\end{itemize}

We assume that once registered to a (primary) owner, a~\ac{TD} cannot be registered to a new (primary) owner.
As we detail in~\sect{sec:securedesign}, we allow primary owners to register secondary owners.

\vspace{-0.1cm}
\subsection{\ac{CD} security properties}

\mypar{Anonymous and proximity-aware location update ($CD-S_{1}$).} A~\ac{CD} should be able to~\emph{anonymously} update the location of~\emph{only} in proximity~\acp{TD}.
This ensures that participating in the crowdsourced system does not pose any privacy risks for the~\ac{CD} and protects against Sybil attacks~\cite{wang2016defending} by ensuring that location can be updated for only those~\acp{TD} that are in BLE proximity 
This constrains that all the sybils need to be in BLE proximity of the victim~\ac{TD}.
To guarantee this property, we require that a~\ac{CD} communicates with a~\ac{TD} to update its location.
Since ~\acp{TD} can communicate only through~\ac{BLE}, such communication ensures proximity.
Hence, the~\ac{NSC} is:
\begin{itemize}
    \item \textbf{\capfive{}:} The~\ac{CD} can update the location of a~\ac{TD} if and only if the~\ac{CD} can communicate with the~\ac{TD}.
\end{itemize}

\vspace{-0.1cm}
\subsection{\ac{TS} security properties}

\mypar{Reliable location service ($TS-S_{1}$).} The location information provided by the \ac{TS}  must be reliable, \ie{}, the retrieved location of a~\ac{TD} must not be spoofable.
This ensures that the location information about a~\ac{TD} provided by the~\ac{TS} can be trusted by the owners.

To have the $TS-S_{1}$ property, all the entities involved in the location update of a~\ac{TD} (\ie{},~\ac{TD}, location information from~\ac{CD}, and, \ac{TS}) should not be affected.
Which results in the following~\acp{NSC}:
\begin{itemize}
    \item \textbf{\capsix{}:} It must be impossible to spoof a given~\ac{TD}.
    \item \textbf{\capseven{}:} It must be impossible to spoof a location by a~\ac{CD}.
    \item \textbf{\capeight{}:} It must be impossible to spoof the~\ac{TS}\footnote{This can be  achieved by ensuring that~\ac{TS} uses~\ac{TLS}~\cite{dierks2008transport}.}.
\end{itemize}

\noindent
\tbl{tbl:securityproperties} summarizes the aforementioned security properties along with the corresponding necessary conditions (\capone{}-\capeight{}), which form the~\acp{NSC} of a secure crowdsourced tracking system.

\begin{table*}[t]
\centering
\scriptsize
\caption{Security properties of an ideal crowdsourced tracking system along with the corresponding necessary conditions.}
\label{tbl:securityproperties}
\vspace{-0.3cm}
\begin{tabularx}{\textwidth}{L{1.3cm} |c c c c}
\toprule
& \multicolumn{4}{c}{\textbf{Security properties of a crowdsource tracking system}} \\
\\
& \multicolumn{2}{c}{\textbf{Tracking device (\ac{TD})}} & \textbf{Communication device (\ac{CD})} & \textbf{Tracking service (\ac{TS})} \\
\cmidrule{2-5}
& \textbf{Privileged owners} & \textbf{Physical ownership} & \textbf{Anonymous \& proximity-aware} & \textbf{Reliable location service} \\
& \textbf{($TD-S_{1}$)} & \textbf{($TD-S_{2}$)} & \textbf{location update ($CD-S_{1}$)} & \textbf{($TS-S_{1}$)} \\

\midrule

\multirow{2}{*}{
    \makecell{
        \textbf{Necessary} \\ \textbf{Conditions}
    }
} 

& 

\makecell[{{p{3.3cm}}}]{
    \tabitem TD should recognize its owners (\capone{}) \\
    \tabitem TS should recognize owners of a given TD (\captwo{})
}

& 

\makecell[{{p{3.3cm}}}]{
    \tabitem Owning a TD should involve performing physical action on it (\capthr{}) \\
    \tabitem TS should be able to verify proof of physical activity (\capfour{})
}

&

\makecell[{{p{3.95cm}}}]{
    \tabitem \ac{CD} can update the location of a~\ac{TD} if and only if the~\ac{CD} can communicate with the~\ac{TD} (\capfive{})
}

&

\makecell[{{p{3.95cm}}}]{
    \tabitem Impossible to spoof a TD (\capsix{}) \\
    \tabitem Impossible to spoof a location by CD (\capseven{}) \\
    \tabitem Impossible to spoof the TS (\capeight{})
}

\\

\bottomrule
\end{tabularx}
\end{table*}

\section{Threat model}
In any crowdsourced tracking system, the~\ac{TS} plays a major role as it is the main entity that receives and provides location information for all the~\acp{TD}. 
In our work, we consider~\ac{TS} to have~\emph{malicious but cautious} behavior~\cite{cheval2014tests}. In practice, on the one hand, we assume that the~\ac{TS} does not intentionally attempt to spoof the location information, as users could easily identify tampered location records.
On the other hand, we assume that the~\ac{TS} can actively attempt to collect or learn information about the location of the~\acp{TD} from incoming reports and queries, therefore affecting the users' privacy.

We assume an external attacker that has full control of the software that runs on a~\ac{CD}, except for the code that is protected through hardware isolation mechanisms, such as ARM TrustZone~\cite{pinto2019demystifying}. 

We consider the \acp{TD} to be a \emph{black box} for the attacker. Specifically, the attacker can only communicate with a~\ac{TD} through~\ac{BLE}.
As such, attacks involving physically tampering with the~\acp{TD} and corresponding hardware attacks are out-of-scope for our work.
However, we assume that the attacker can snoop in the~\ac{BLE} traffic and try to mimic a~\ac{TD}.
Finally, we do not consider vulnerabilities in communication protocols implementations such as~\ac{BLE} and TLS.

Similarly, software vulnerabilities in the code that runs in the~\ac{TS} and~\acp{TD} are out-of-scope, and well covered by other research work.
However, we assume that the~\ac{TS} can be subject to data breaches, and thus we aim at protecting location information even in such scenarios---guaranteeing end-to-end encryption.

In our threat model, attackers are interested in accessing or spoofing information about the location of users' valuable objects, with the goal of, for instance, tracking and stealing such valuables.

%
%
%

\section{Security analysis}
We study five popular crowdsourced tracking systems, TrackR~\cite{trackr}, CUBE~\cite{cubetr}, Chipolo~\cite{chipolotr}, Pebblebee~\cite{pebblebeetr}, and, Tile~\cite{tile}, looking at whether each system guarantees the aforementioned security properties (\sect{subsub:securityprop}).
We do so by verifying if all the~\acp{NSC} hold.
All of the studied systems provide a~\ac{TApp} that users can use to communicate with the~\acp{TD} and the corresponding~\ac{TS} (\tbl{tbl:crowdtrackingdataset}).
Users must first create an account by registering with the~\ac{TS}.
All communications with the~\ac{TS} then require the user to login and include their session token in every request sent to the~\ac{TS}.
We use the term~\emph{authenticated REST endpoint} to indicate that the endpoint requires the user to authenticate first.
Some tracking services allow a~\ac{TD} to have multiple owners so that multiple users can share and track a~\ac{TD}.
Note that, there exist other~\ac{BLE} tracking systems, such as GoFinder HL~\cite{gofindertr} and Nut~\cite{nuttkr}, which, however, do not adopt the crowdsourced model and, hence, are out-of-scope for this study. 

\subsection{Methodology}

Our analysis methodology is based on the following two steps.

\mypar{Record.}
We record the messages sent from the~\ac{TApp} to the~\ac{TD} and~\ac{TS}.
First, we manually reverse engineer the~\ac{TApp} to identify the functions used to send and receive messages from and to the~\ac{TS} and~\ac{TD}.
For instance, the TrackR's~\ac{TApp} uses the~\texttt{write} method of the \texttt{java.io.BufferedWriter} class to send messages to the~\ac{TS}.
Next, we setup a dynamic instrumentation environment to hook into the identified functions and record all their messages.
Finally, we use the corresponding~\ac{TApp} to register a~\ac{TD}, update its location from a different \ac{CD}, and retrieve the updated location.
Our dynamic instrumentation records all the messages sent through the instrumented functions.
We carefully select the appropriate functions to hook so that we get messages in plaintext before they encrypted~\cite{chen2018iotfuzzer}.
Thus, hooking at lower layers (\eg, system calls) might not reveal the content of the messages.

We report the details of the communication endpoints of each tracking system in the Appendix of our full version~\cite{secrowfull}.

\mypar{Verify.}
\label{subsub:methverify}
We use recorded messages to test the required security properties.
For instance, consider the case of verifying the physical ownership property for CUBE.
We purchased two CUBE~\acp{TD}, $D1$, and $D2$, whose Bluetooth MAC address is respectively \texttt{MAC\_1} and \texttt{MAC\_2}. 
During our recording phase, we record that, when registering a new owner for $D1$, the~\ac{TApp} sends an HTTP POST request to the endpoint \texttt{/api/devices?AddCubeWithMac=MAC\_1} with the following JSON payload: \texttt{\{"name":"<NAME>", "mac": "MAC\_1"\}}.

We can see that the payload includes the MAC address of $D1$ (\ie{}, \texttt{MAC\_1}) and we do not see any communication with $D1$, which violates~\ac{NSC} $\capfour{}$.
Now, to verify the physical ownership in CUBE, we replay the request by replacing the MAC address with \texttt{MAC\_2}.
A successful request indicates that we can register a new owner for the~\ac{TD} without physical access, thus violating the physical ownership property ($TD-S_{2}$).
We follow the same black-box methodology to verify each property for each tracking system.


In the following sections, we describe the results of our analysis of the studied crowdsourced tracking systems, and we show that none of them guarantee the desired security properties.

\begin{table}[b]
\footnotesize
\setlength{\tabcolsep}{3pt}
\caption{Crowdsourced tracking systems considered in our security analysis.}
\label{tbl:crowdtrackingdataset}
\vspace{-0.3cm}
\resizebox{\columnwidth}{!}{

    \begin{tabular}{l c c c}
    \toprule
    \textbf{Tracking} & & & \textbf{Multi-owner}\\
    \textbf{System} & \textbf{\ac{TApp}} & \textbf{\ac{TS}} & \textbf{Support}\\
    \midrule
    \rowcolor{black!15} TrackR~\cite{trackr}          &  com.phonehalo.itemtracker~\cite{trackrapp}  &  \url{platform.thetrackr.com} & Yes  \\
    CUBE~\cite{cubetr}            &  com.blueskyhomesales.cube~\cite{cubeapp}  &  \url{net.cubetracker.com}  & No\\
    \rowcolor{black!15} Chipolo~\cite{chipolotr}         & chipolo.net.v3~\cite{chipoloapp}   &  \url{api.chipolo.net}  & No\\
    Pebblebee~\cite{pebblebeetr}       &  com.pebblebee.app.hive3~\cite{pebblebeeapp}  & \url{api.pebblebee.com} & No  \\
    \rowcolor{black!15} Tile~\cite{tile}          &  com.thetileapp.tile ~\cite{tileapp}  &  \url{production.tile-api.com} & No  \\
    \bottomrule
    \end{tabular}

}
\end{table}

\subsection{TrackR}



\mypar{Owner registration.} 
\label{subsub:trackraddowners}
A user can own a~\ac{TD} by pairing with it 
through BLE. 
Furthermore, the user also needs to send a \texttt{POST} request to~\ac{TS} to register as an owner of the~\ac{TD}.
Specifically, the registration request is sent to the REST endpoint with a JSON payload containing a unique identifier, called~\texttt{trackerid}, of the~\ac{TD} along with ~\texttt{USERTOKEN}.
Here,~\texttt{USERTOKEN} is the login token of the user, and the~\texttt{trackerid} of the~\ac{TD} is a string composed of four zeros (``0000") and the reverse of the Bluetooth MAC address of the~\ac{TD}.
For instance, for a~\ac{TD} with MAC address 00:1B:44:11:3A:B7, the~\texttt{trackerid} is 0000b73a-11441b00.

This design \emph{violates} both the \emph{\textbf{Privileged owners ($TD-S_{1}$)}} and the \emph{\textbf{Physical ownership ($TD-S_{2}$)}} properties.
In fact, the~\ac{TD}, in the case of TrackR, does not keep track of the owners resulting in missing~\capone{}~\ac{NSC}.
Because of this, an attacker can perform privileged operations, such as ringing and blink, on the~\ac{TD} by sending messages over~\ac{BLE} to the~\ac{TD}.
Second, as mentioned above, registering as an owner of a~\ac{TD} requires an attacker to send a POST request with the attacker's login token and the~\texttt{trackerid} of the~\ac{TD}.
However, the~\texttt{trackerid} of a~\ac{TD} can be easily obtained by observing its MAC address, which is broadcasted over Bluetooth and can be known without physical access to the~\ac{TD}.
Thus an attacker can register as an owner of any nearby~\ac{TD}, which violates missing~\capthr{} and~\capfour{}~\acp{NSC}.
Moreover, given that the first three bytes are unique for an organization~\cite{heydon2012bluetooth}, \ie, TrackR, by knowing a MAC address~\emph{an attacker can use the last three bytes to enumerate all possible TrackR~\acp{TD} MAC addresses and register as an owner for every TrackR~\ac{TD}} without having physical access to the~\acp{TD}, thus violating~\emph{Physical ownership} property.

\vspace{-0.1cm}
\mypar{Location update.}
\label{subsub:trackrlocupdate}
Updating the location of a~\ac{TD} to the~\ac{TS} is done by sending a PUT request to the REST interface with a JSON payload containing the~\texttt{trackerid} of the~\ac{TD} along with location information in the form of longitude and latitude values.

This allows an attacker to update the location information of any~\ac{TD} (given its MAC address) without being physically close to the~\ac{TD}.
Note that, updating the location of~\ac{TD} does not require communicating with the~\ac{TD}, consequently missing~\capfive{}~\ac{NSC}.
This~\emph{violates the \textbf{Anonymous and proximity-aware location update ($CD-S_{1}$)} property}.

\vspace{-0.1cm}
\mypar{Location query.}
A user can obtain the location of a~\ac{TD} by sending a GET request to the REST interface using the login token of the user. 
The~\ac{TS} responds with location information of all the~\acp{TD} owned by the user.
Given a~\ac{TD}, an attacker can affect the~\ac{TD}'s location information either by updating its location using the method described in~\sect{subsub:trackrlocupdate} or~\emph{by spoofing the MAC address of the~\ac{TD}} (\ie{}, missing~\capsix{}~\ac{NSC}).
Moreover, the~\ac{TS} blindly trusts the location information about a~\ac{TD} provided by any user without verifying the accuracy of the location information missing~\capseven{}~\ac{NSC}.
These attacks clearly~\emph{violate the \textbf{Reliable location service ($TS-S_{1}$)} property}.

\subsection{CUBE}

\mypar{Owner registration.}
A new owner can register by pairing with the~\ac{TD} and sending a POST request to the~\ac{TS}, through an authenticated REST endpoint. The payload of this request contains the MAC address of the~\ac{TD} along with other data items.

The~\ac{TS} allows a single owner to a given~\ac{TD}. Hence, the~\ac{TS} allows a user to register as the owner of a~\ac{TD} only if the~\ac{TD} is not already registered by another user. However, the~\ac{TS} does not verify that the user has physical access to the~\ac{TD}. This enables an attacker to own an unregistered~\ac{TD} by knowing its MAC address. 
This ~\emph{violates the~\textbf{Physical ownership ($TD-S_{2}$)} property}. Furthermore, as mentioned in~\sect{subsub:trackraddowners}, from a single MAC address of a~\ac{CD} an attacker can enumerate all the possible MAC address of the~\acp{TD} and register as the owner before a valid user attempts to register. 

Similarly to TrackR, the~\acp{TD} in CUBE do not keep track of their owner (\ie{}, missing~\capone{}). This allows an attacker to perform privileged operations, such as ringing and blink, on the~\ac{TD}, thus~\emph{violating the~\textbf{Privileged owners ($TD-S_{1}$)} property}.

\mypar{Location update.}
\label{subsub:cubelocupdate}
In CUBE, location updates by a user are accepted for a~\acp{TD} only if either the~\ac{TD} is owned by the same user or the~\ac{TD} is marked as lost by their respective owners.
The location update of a~\ac{TD} is sent to the~\ac{TS} using a PUT request with a JSON body containing the device MAC address and location information.

Similar to TrackR, the CUBE~\ac{TS} blindly trust the location information about a~\ac{TD} provided by any user without verifying neither the accuracy of the location information (\ie{}, missing~\capseven{}) not the physical proximity of the~\ac{TD} (\ie{} missing~\capfive{}).
This~\emph{violates the~\textbf{Anonymous and proximity-aware location update ($CD-S_{1}$)} property} as an attacker can update and spoof the location of any~\ac{TD} without being physically close to them.

\mypar{Location query.}
The location of a~\ac{TD} can be obtained by sending a GET request along with the emailID of the user account.
This returns the location of all the~\acp{TD} registered by the user.

Similar to TrackR, in the case of CUBE, an attacker can affect the~\ac{TD}'s location information either by updating the location using the method described in~\sect{subsub:cubelocupdate} or~\emph{by spoofing MAC address of the~\ac{TD}} (\ie{}, missing~\capsix{}).
These attacks clearly~\emph{violate the \textbf{Reliable location service ($TS-S_{1}$)} property}.

\subsection{Chipolo}
In Chipolo, the~\ac{TS} assigns to every~\ac{TD} and user a unique id, called \texttt{chipoloid}, which is used by the users to communicate with the~\ac{TS} on behalf of the~\ac{TD}.

\mypar{Owner registration.}
In Chipolo, similar to other tracking systems, a~\ac{TD} can be owned by first pairing with it.
However, registering the ownership of a~\ac{TD} to the~\ac{TS} requires two REST calls.
To register as the owner for a~\ac{TD}, first, the user needs to send a GET request containing the unique id assigned to the user on account registration and the Bluetooth MAC address of the~\ac{TD}.

If the~\ac{TD} is not registered~\ac{TS} sends a one-time secret device token. 
Next, the user needs to make a POST request with JSON payload containing the above secret token. 
On successful registration,~\ac{TS} sends a JSON response with a success message containing the~\texttt{chipoloid} that can be used for all future communication with~\ac{TS} concerning the~\ac{TD}.
However, similar to other tracking systems, the~\ac{TS} does not verify that the~\ac{CD} can communicate with the~\ac{TD} (\ie{}, missing~\captwo{}).
This enables an attacker to own an unregistered~\ac{TD} by only knowing its MAC address, which does not require physical access.
This ~\emph{violates the~\textbf{Physical ownership ($TD-S_{2}$)} property}

Furthermore, unlike other tracking systems, only owners~\ie{}, paired users, can send privilege commands (using a~\ac{CD}) to a~\acp{TD}, thus \emph{preserving the~\textbf{Privileged owners ($TD-S_{1}$)} property}.

\mypar{Location update.}
\label{subsub:chipololocupdate}
Similarly to owner registration, updating the location of a~\ac{TD} requires two REST calls.
To update the location of a~\ac{TD}, first, a GET request should be sent with the unique id for the user and~\texttt{DEVICEID}, which is an identifier broadcasted by the~\ac{TD} through one of the advertised packets.

This GET request returns the~\texttt{chipoloid}, which is then used in the second PUT request to the authenticated REST endpoint, with the payload containing the location information.

Similar to other systems, the~\ac{TS} of Chipolo blindly trusts the location information about a~\ac{TD} provided by any user without verifying neither the accuracy of the location information (\ie{}, missing~\capseven{}) nor the physical proximity of the~\ac{TD} (\ie{}, missing~\capfive{}).
Consequently, this~\emph{violates the \textbf{Anonymous and proximity-aware location update ($CD-S_{1}$)} property}.

\mypar{Location query.}
A user can obtain the location information of all the~\acp{TD} by sending a GET request, which returns a JSON response containing the location information of all the~\acp{TD} owned by the user corresponding to~\texttt{USERID}.

As explained in~\sect{subsub:chipololocupdate}, given the MAC address of a~\ac{TD},
an attacker can update the location of the~\ac{TD}, thereby compromising the integrity of location information.
Thus~\emph{violating the \textbf{Reliable location service ($TS-S_{1}$)} property}.

\subsection{Pebblebee}
\vspace{-0.1cm}
\mypar{Owner registration.}
To register as the owner of a~\ac{TD}, a user needs to first pair with the~\ac{TD} through~\ac{BLE}.
To register with the~\ac{TS}, a POST request needs to be sent with the Bluetooth MAC address of the~\ac{TD}. 
During registration, the~\ac{TS} does not verify that the user can communicate with the~\ac{TD}, consequently missing~\capfour{}. 
This enables an attacker to own an unregistered~\ac{TD} by knowing its MAC address, which does not require physical access, and, therefore, \emph{violates the~\textbf{Physical ownership ($TD-S_{2}$)} property}.

Similar to TrackR and CUBE, the~\ac{TD} in the case of Pebblebee does not keep track of the owner (\ie{}, missing~\capone{}) which enables an attacker to perform privileged operations, such as ringing and blink, on the~\ac{TD} by sending messages over~\ac{BLE}, thus~\emph{violating the~\textbf{Privileged owners ($TD-S_{1}$)} property}.

\mypar{Location update.}
\label{subsub:pebblelocupdate}
In Pebblebee, location update of a~\ac{TD} is done by sending a POST request with a JSON payload containing the location information along with the MAC address of the~\ac{TD}.

Similar to other systems, the~\ac{TS} of Pebblebee does not verify either the accuracy of the location information (\ie{}, missing~\capseven{}) nor the physical proximity of the~\ac{TD} (\ie{}, missing~\capfive{}).
Consequently, this~\emph{violates the \textbf{Anonymous and proximity-aware location update ($CD-S_{1}$)} property}.

\mypar{Location query.}
The location query in Pebblebee is done by sending a GET request, with the Bluetooth MAC address of the~\ac{TD}.
Unfortunately, given the~\texttt{MAC ADDRESS} of a~\ac{TD},~\emph{any user} can query the location (a privileged operation) of the~\ac{TD} using the above URL.
Furthermore, as explained in~\sect{subsub:trackraddowners}, from a single MAC address of a~\ac{TD}, it is easy to enumerate the MAC address of all~\acp{TD}.
Consequently, an attacker can perform location query of all~\acp{TD}, which further~\emph{violates the~\textbf{Privileged owners ($TD-S_{1}$)} property}.

Given that an attacker can arbitrary update the location of any~\ac{TD} (\sect{subsub:pebblelocupdate}), this affects the integrity of location information.
Thus~\emph{violating the \textbf{Reliable location service ($TS-S_{1}$)} property}.

\subsection{Tile}

\mypar{Owner registration.}
\label{subsub:tileownerreg}
Similar to the other systems,~\ac{TD} in Tile needs to be first paired through BLE using~\ac{TApp},
which will then register with ~\ac{TS} by sending a PUT request to it's API interface at~\url{https://production.tile-api.com/api/v1/}.
Along with this request, a unique signature and static client uuid are added. 
This requires the~\ac{TD} to be physically present near the~\ac{CD}. 
Unlike other systems, once a Tile is added it cannot be deleted by the owner. 
The owner, however does have other options such as to hide~\ac{TD} temporarily, 
transfer it to another user using their email address and replace an existing Tile with a new Tile. 
The owner registration process of Tile is the most secure design of all the analyzed systems.
In fact, our design as explained in~\sect{subsec:ownerregis} is inspired by Tile.

\mypar{Location update}
\label{subsub:tilelocupdate}
In Tile, location updates are encrypted and updated by sending a POST request to the~\ac{TS}.
But, only logged in users can send the location update request thereby~\emph{violating the \textbf{Anonymous and proximity-aware location update ($CD-S_{1}$)} property}.

\mypar{Location query}
Location query in Tile is done by sending a authenticated GET request, with requesting client uuid, and a tile request signature of the~\ac{CD}. 
Furthermore, only owner can query the location of a~\ac{TD}.

Finally, given that the location information is blindly trusted. An attacker
can update all in-proximity~\acp{TD}s with random locations thereby~\emph{violating the \textbf{Reliable location service ($TS-S_{1}$)} property}.
\begin{table*}[t]
    \centering
    \scriptsize
    \caption{Verified security properties for each device, where \ding{51} and \ding{55} in each cell indicate whether the corresponding security property holds or not respectively. The~\acp{NSC} ($C_{x}$) under \ding{55} indicates the~\emph{missing conditions (\sect{subsub:securityprop})}.}
    \label{tab:verifiedsecurityproperties}
    \vspace{-0.3cm}
    \begin{tabularx}{0.9\textwidth}{X | c c c c }
    \toprule
    & \multicolumn{4}{c}{\textbf{Security Properties of a crowdsourced tracking system}} \\ 
    & \multicolumn{2}{c }{\textbf{Tracking device (TD)}} & \textbf{Communication device (CD)} & \textbf{Tracking service (TS)} \\
    \midrule
    \textbf{Device manufacturer} & 
    \makecell{ \textbf{Privileged owners}\\ \textbf{($TD-S_{1}$)} } &
    \makecell{ \textbf{Physical ownership}\\ \textbf{($TD-S_{2}$)} } &
    \makecell{ \textbf{Anonymous and proximity-aware location update}\\ \textbf{($CD-S_{1}$)}} &
    \makecell{ \textbf{Reliable location service}\\ \textbf{($TS-S_{1}$)} } \\
    \midrule

    \rowcolor{black!15} TrackR & \ding{55} (\capone{}) & \ding{55} (\capthr{}, \capfour{}) & \ding{55} (\capfive{}) & \ding{55} (\capsix{}, \capseven{}) \\ 
    CUBE & \ding{55} (\capone{}) & \ding{55} (\capthr{}, \capfour{}) & \ding{55} (\capfive{}) & \ding{55} (\capsix{}, \capseven{}) \\ 
    \rowcolor{black!15} Chipolo & \ding{51} & \ding{55} (\capthr{}, \capfour{}) & \ding{55} (\capfive{}) & \ding{55} (\capseven{}) \\ 
    Pebblebee & \ding{55} (\capone{}) & \ding{55} (\capthr{}, \capfour{}) & \ding{55} (\capfive{}) & \ding{55} (\capsix{}, \capseven{}) \\ 
    \rowcolor{black!15} Tile & \ding{51} & \ding{51} & \ding{55} (\capfive{}) & \ding{55} (\capseven{}) \\
    \bottomrule
    \end{tabularx}
\end{table*}
\vspace{-0.1cm}
\subsection{Summary}
\label{subsec:analysissum}
\tbl{tab:verifiedsecurityproperties} summarizes the results of our analysis. 
We verified all our findings by implementing a~\ac{PoC} for each violation.
Interestingly, all the tacking systems satisfy two~\acp{NSC}, \ie,~\captwo{} (\ac{TS} able to recognize owners of a~\ac{CD}) and~\capeight{} (Impossible to spoof the~\ac{TS}), which they achieve by using HTTPS for their communications.
Our results show that, except for the \textbf{Privileged owners ($TD-S_{1}$)} property in Chipolo, \emph{none of analyzed tracking systems guarantee the desired security properties}.

\mypar{Responsible disclosure.}
All the device vendors have been contacted and notified regarding the security issues discussed in the paper. 
Two vendors already replied and acknowledged our findings.
We have not heard back from the remaining two vendors yet.

\vspace{-0.2cm}
\section{Secure tracking service}
\label{sec:securedesign}

In this section, we present~\systemname{}, a crowdsourced tracking system that satisfies all the desired security properties.
Similar to the existing systems, the~\ac{TS} in~\systemname{} requires any user to register and log in before using the system.
Furthermore, the~\ac{TS} keeps track of the ownership association~\ie{}, the set of owners of a~\ac{TD}, and is also able to recognize an owner of a~\ac{TD}. This satisfies our~\ac{NSC}~\captwo{}.
Furthermore, the~\ac{TS} uses~\ac{TLS} for all its communication, thereby making it impossible to spoof a~\ac{TS}, which satisfies~\ac{NSC}~\capeight{}.

\vspace{-0.15cm}
\subsection{Public key cryptography}
\label{subsub:cryptoassumptions}
One of the main building blocks of~\systemname{} is the public key cryptography~\cite{salomaa2013public}.
Specifically, every entity in our system,~\ie{},~\ac{TS},~\ac{CD}, and,~\ac{TD}, has a public key and private key pair.
As in all public key cryptographic systems, the public key of an entity is available to the other entities through certificates.
The private key is, instead, secret and only known to the corresponding entity.

\mypar{Notations.}
We use the following notations for various cryptographic primitives.
$Pk_{x}$ and $Pr_{x}$ indicate public and private key of entity $x$.
$Enc(K, P)$ and $Sign(R, P)$ indicates encryption of content $P$ with~\emph{public key} $K$ and signature of $P$ with $R$, in the case of RSA this is same as encryption of $P$ with~\emph{private key} $R$.
Since encryption with the private key is called a signature, we use $Sign$ to indicate encryption with a private key.
Also, $SEnc(S, P)$ indicates encryption with the~\emph{symmetric key} $S$
$X_{I}$ indicates a unique~\emph{identifier} (ID) for the entity $X$.
For instance, this could be the Bluetooth MAC address for a~\ac{TD}, and the user $ID$ of the logged-in user
for a~\ac{CD}.

\subsection{Built in capabilities}
\label{subsec:builtincapabilities}
In~\systemname{}, we assume the~\acp{TD} and~\acp{CD} to have certain capabilities, which allow them to provide the required security properties.

\mypar{\ac{TD} capabilities.}
\label{subsubsec:tdcapabilities}
The~\ac{TD} has its private key etched on to on-chip read-only memory, such as e-fuses~\cite{erickson2017implementing}. Thus, to spoof a given~\ac{TD}, one needs access to its private key, which is impossible without physically invasive techniques~\cite{lohrke2018key}.
The~\ac{TD} also has a small amount of persistent storage and the ability to generate random numbers.
The persistent storage contains the~\emph{location key (\ie{}, $L_{\ac{TD}}$)}, which is used as the~\emph{symmetric key} to encrypt and decrypt location data of the~\ac{TD} (\encircle{4} in~\fig{fig:seclocupdate}).
The persistent storage also contains the public key certificates of all the owners of the~\ac{TD}.
This enables a~\ac{TD} to verify whether a given~\ac{CD} is one of its owners and thus to satisfy the~\ac{NSC}~\capone{}.
Every~\ac{TD} has a~\emph{single} primary owner and can have multiple secondary owners.
The primary owner has additional privileges and can regulate the secondary owners of the~\ac{TD}.

\mypar{\ac{CD} capabilities.}
\label{subsubsec:cdcap}
All the location updates in~\systemname{} have to be signed.
Since~\acp{CD} (\ie{}, smartphones) are commonly used to communicate with the~\ac{TS},~\acp{CD} have to be able to provide~\emph{signed location (GPS) information}.
Furthermore, the~\ac{CD} operating system must not be able to tamper the location information.
This can be achieved by using existing techniques~\cite{li2018vbu} that leverage~\acp{TEE}, such as ARM TrustZone~\cite{pinto2019demystifying}, which is now available across almost all the smartphone brands~\cite{cerdeira2020sok}.
Furthermore,~\acp{CD} have the ability to create temporary but attested public-private key pairs, denoted as $TPk_{CD}$ and $TPr_{CD}$, which are managed by a~\ac{TEE} and cannot be tampered with by the user operating system~\cite{androidkeystore, applekeystore}.
As we show in~\sect{subsub:locupdate}, these keys are used to provide anonymous but signed location information.
In this scheme, the~\acp{CD} cannot provide fake location information, therefore satisfying one of the~\acp{NSC},~\ie{},~\capseven{}.

\subsection{Owner registration}
\label{subsec:ownerregis}
\begin{figure}[t]
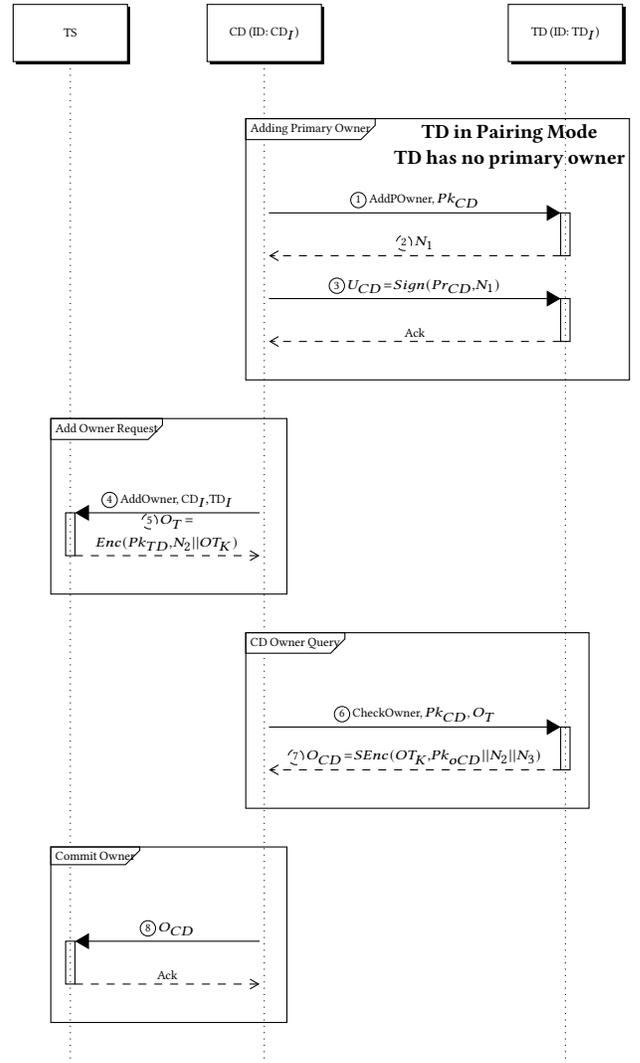

    \centering
    \tiny

    \tikzset{
    every picture/.append style={
        transform shape,
        scale=0.95
    }
    }

    \begin{sequencediagram}
    \newinst{TS}{\ac{TS}}
    \newinst[1.1]{CD}{\ac{CD} (ID: $\ac{CD}_{I}$)}
    \newinst[2.6]{TD}{\ac{TD} (ID: $\ac{TD}_{I}$)}

    \begin{sdblock}{Adding Primary Owner}{\shortstack{\textbf{\ac{TD} in Pairing Mode}\\ \textbf{\ac{TD} has no primary owner}}}
    \begin{call}
    {CD}{\encircle{1} AddPOwner, $Pk_{CD}$}
    {TD}{\encircle{2} $N_{1}$}
    \end{call}
    \begin{call}
    {CD}{\shortstack{\encircle{3} $U_{CD}=Sign(Pr_{CD}, N_{1})$}}
    {TD}{Ack}
    \end{call}
    \end{sdblock}

    \begin{sdblock}{Add Owner Request}{}
    \begin{call}
    {CD}{\encircle{4} AddOwner, $\ac{CD}_{I}, \ac{TD}_{I}$}
    {TS}{\shortstack{\encircle{5} $O_{T}=$\\$Enc(Pk_{TD}, N_{2}||OT_{K})$}}
    \end{call}
    \end{sdblock}

    \begin{sdblock}{CD Owner Query}{}
    \begin{call}
    {CD}{\encircle{6} CheckOwner, $Pk_{CD}$, $O_{T}$}
    {TD}{\shortstack{\encircle{7} $O_{CD}=SEnc(OT_{K}, Pk_{oCD}||N_{2}||N_{3})$}}
    \end{call}
    \end{sdblock}

    \begin{sdblock}{Commit Owner}{}
    \begin{call}
    {CD}{\encircle{8} $O_{CD}$}
    {TS}{Ack}
    \end{call}
    \end{sdblock}

    \end{sequencediagram}
\vspace{-0.3cm}
\caption{Secure owner registration of ~\systemname{}}
\vspace{-0.4cm}
\label{fig:secownerregister}
\end{figure}

In~\systemname{}, adding owners to a~\ac{TD} requires the~\ac{TD} to be in pairing mode, which is only possible by physically holding down a dedicated button on the~\ac{TD}. This satisfies~\capthr{}~\ac{NSC} (\sect{subsub:tdproperties}).
\fig{fig:secownerregister} shows the sequence diagram of the protocol used in~\systemname{} to add the primary owner to a~\ac{TD} and the~\ac{TS}.

The first block~\emph{Adding Primary Owner} shows the messages exchanged when a~\ac{CD} (or user) requests to be added as the owner of the~\ac{TD}.
When a~\texttt{AddPOwner} request along with the public key, $Pk_{CD}$, is received by a~\ac{TD}, the~\ac{TD} sends a challenge nonce $N_{1}$ back to the~\ac{CD}. The~\ac{CD} is expected to sign the nonce $N_{1}$ to prove that it has access to the private key of the corresponding public key sent in the initial message. On successful verification of the signature by the~\ac{TD}, the~\ac{TD} adds the pubic key~$Pk_{CD}$ as the primary owner.
The~\texttt{AddPOwner} request is accepted by a~\ac{TD}~\emph{only when there is no primary owner yet} in the~\ac{TD}.
The primary owner can add secondary owners using the~\texttt{AddSOwner} command (\sect{subsubsec:primaryownercmds}).

Registering as an owner (either primary or secondary) of a~\ac{TD} to the~\ac{TS} requires three interactions, as indicated by the bottom three blocks in~\fig{fig:secownerregister}. 
First, the~\ac{CD} sends a~\texttt{AddOwner} request along with its ID ($CD_{I}$) and the ID of the~\ac{TD} ($TD_{I}$) to the~\ac{TS}.
The~\ac{TS} replies with a nonce $N_{2}$ and a temporary symmetric key $OT_{K}$ both encrypted with the public key of~$\ac{TD}_{I}$ (\ie{}, $O_{T}$).
Second (\texttt{CD Owner Query}), the~\ac{CD} relays $O_{T}$ along with its public key $Pk_{CD}$ to the~\ac{TD}.
The~\ac{TD} decrypts $O_{T}$ (as it has the private key $Pr_{\ac{TD}}$) to retrieve $N_{2}$ and the temporary~\emph{symmetric key} $OT_{K}$.
Using this temporary key, $OT_{K}$, the~\ac{TD} sends back an encrypted message ($O_{CD}$) that includes the public key of the owner ($Pk_{oCD}$), the decrypted nonce ($N_{2}$), and, an additional nonce for freshness ($N_{3}$).
Here, $Pk_{oCD} = Pk_{CD}$, if $Pk_{CD}$ is one of the owners of the~\ac{TD}, else $Pk_{oCD}$ is the public key of one of the registered owners.
The use of $OT_{K}$ for encryption ensures that only the~\ac{TS} can know the owner information of the~\ac{TD}, thereby protecting the privacy of the~\ac{TD}.
Furthermore, the freshness token $N3$ prevents inference attacks.
Finally, in~\texttt{Commit Owner}, the~\ac{CD} relays $O_{CD}$ to the~\ac{TS}.
The~\ac{TS} decrypts $O_{CD}$ as it has access to the private key of~\ac{TD} (\sect{subsub:cryptoassumptions}), retrieves the $Pk_{oCD}$, verifies that the decrypted nonce matches $N_{2}$, and, discards $N_{3}$.
If the $Pk_{oCD}$ is the public key of $\ac{CD}_{I}$ then~\ac{TS} adds $\ac{CD}_{I}$ as one of the owners of the~\ac{TD} (or~$\ac{TD}_{I}$).
The validation of the nonce $N_{2}$, through $O_{CD}$, ensures that the ownership request is approved by the~\ac{TD}, therefore satisfying our~\capfour{}~\ac{NSC}.
 
The protocol described above satisfies the~\textbf{Physical ownership ($\ac{TD}-S_{2}$)} property, \ie{}, it is impossible to own a tracking device without physical access.
This is because, first, adding an owner to a~\ac{TD} requires the~\ac{TD} to be in pairing mode, which is only possible by physically pressing a button on the~\ac{TD}.
Second, to register to the~\ac{TS}, the~\ac{CD} must be able to provide \emph{a valid $O_{CD}$}, \ie, the public key of the owner and correct nonce ($N_{2}$), encrypted with the temporary symmetric key $OT_{K}$.
The~\ac{CD} cannot produce a valid $O_{CD}$ because it requires access to $N_{2}$ and $OT_{K}$, but, both are encrypted ($O_{T}$) with the public key of~\ac{TD}.
However, generating a valid $O_{CD}$ from $O_{T}$ is possible by communicating with the~\ac{TD} (\ie{} \texttt{CD Owner Query} of~\fig{fig:secownerregister}).
Note that, the~\ac{TD} generates a valid $O_{CD}$ containing the public key of the~\ac{CD} if and only if the~\ac{CD} is one of its owners. As mentioned before, adding owners requires physical access.
Instead, if the~\ac{CD} is not an owner of the~\ac{TD}, the~\ac{TD} still produces a valid $O_{T}$, but containing the public key of a real owner.

\mypar{Primary owner commands.}
\label{subsubsec:primaryownercmds}
As mentioned in~\sect{subsubsec:tdcapabilities}, the primary owner of a~\ac{TD} has additional privileges.
Specifically, primary owners can issue the following commands:
(1) \texttt{UpdateLocKey}  to update the location key (\ie{}, $L_{\ac{TD}}$) of the~\ac{TD}; (2) \texttt{AddSOwner} \& \texttt{RemSOwner} to add or remove secondary owners of the~\ac{TD}.

\begin{figure}[t]
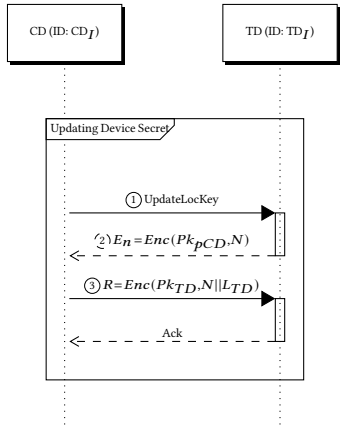

    \centering
    \tiny

    \tikzset{
    every picture/.append style={
        transform shape,
        scale=0.95
    }
    }

    \begin{sequencediagram}
    \newinst{CD}{\ac{CD} (ID: $\ac{CD}_{I}$)}
    \newinst[1.4]{TD}{\ac{TD} (ID: $\ac{TD}_{I}$)}

    \begin{sdblock}{Updating Device Secret}{}
    \begin{call}
    {CD}{\encircle{1} UpdateLocKey}
    {TD}{\encircle{2} $E_{n}=Enc(Pk_{pCD}, N)$}
    \end{call}
    \begin{call}
    {CD}{\shortstack{\encircle{3} $R=Enc(Pk_{TD}, N||L_{TD})$}}
    {TD}{Ack}
    \end{call}
    \end{sdblock}

    \end{sequencediagram}
\vspace{-0.3cm}
\caption{Updating the location key of a~\ac{TD} in~\systemname{}}
\vspace{-0.2cm}
\label{fig:updatedevicesecret}
\end{figure}

Each of the above commands involves a challenge nonce issued by the~\ac{TD}, which is encrypted with the stored public key of the primary owner.
The requesting~\ac{CD} can decrypt the message and obtain the nonce only if it is indeed the primary owner.
The decrypted nonce is used to authenticate the following requests.
\fig{fig:updatedevicesecret} shows the flow of the~\texttt{UpdateLocKey} procedure in~\systemname.
Here, when the command is issued by a~\ac{CD} (\encircle{1}), the~\ac{TD} responds (\encircle{2}) back with an encrypted random nonce ($E_{n}$)---encrypted with the stored public key of the primary owner ($Pk_{pCD}$).
The~\ac{CD}, if it is the primary owner (\ie{}, the~\ac{CD} has the corresponding private key), can decrypt $E_{n}$ to retrieve $N$.
The~\ac{CD} now uses $N$ to send the data (\encircle{3}) of the request ($R$) to the~\ac{TD},~\ie, the location key $L_{TD}$ along with $N$, both encrypted with the public key of the~\ac{TD}.
Finally, the~\ac{TD}, after receiving $R$, decrypts and verifies the nonce $N$. On a successful match, the~\ac{TD} stores the provided key, $L_{TD}$, as the location key.
As we show in~\sect{subsec:locquery}, the location key of a~\ac{TD} is required to obtain the latest location of the~\ac{TD}.
We consider that~\emph{primary owner to be responsible for sharing the location key with secondary owners}.
This provides the primary owner an access control mechanism to regulate who (secondary owners) can access the location information of the~\ac{TD}.
Note that, by using a random nonce, our design is not vulnerable to replay attacks.

The commands~\texttt{AddSOwner} and \texttt{RemSOwner(s)} follow a procedure similar to the one shown in~\fig{fig:updatedevicesecret}. However, in the request data $R$, the primary owner sends the public key of the~\ac{CD} to be added or removed instead of the location key.

\vspace{-0.1cm}
\subsection{Location update}
\label{subsub:locupdate}
\begin{figure}[t]
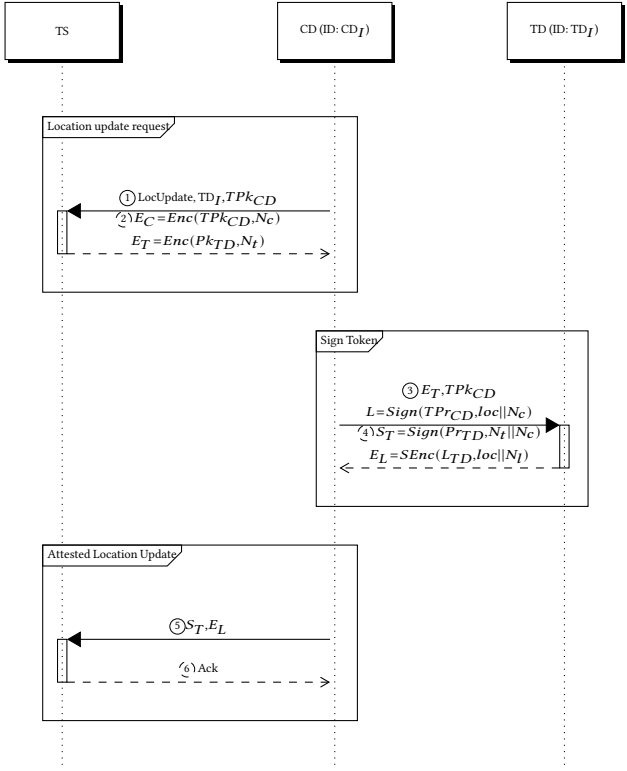

    \centering
    \tiny

    \tikzset{
    every picture/.append style={
        transform shape,
        scale=0.95
    }
    }

    \begin{sequencediagram}
    \newinst{TS}{\ac{TS}}
    \newinst[2.2]{CD}{\ac{CD} (ID: $\ac{CD}_{I}$)}
    \newinst[1.6]{TD}{\ac{TD} (ID: $\ac{TD}_{I}$)}
    \begin{sdblock}{Location update request}{}
    \begin{call}
    {CD}{\encircle{1} LocUpdate, $\ac{TD}_{I}, TPk_{CD}$}
    {TS}{\shortstack{\encircle{2} $E_{C}=Enc(TPk_{CD}, N_{c})$ \\ $E_{T}=Enc(Pk_{TD}, N_{t})$ }}
    \end{call}
    \end{sdblock}
    \begin{sdblock}{Sign Token}{}
    \begin{call}
    {CD}{\shortstack{\encircle{3} $E_{T}, TPk_{CD}$ \\ $L=Sign(TPr_{CD},loc||N_{c})$}}{TD}{\shortstack{\encircle{4} $S_{T}=Sign(Pr_{TD}, N_{t} || N_{c})$ \\ $E_{L} = SEnc(L_{TD}, loc||N_{l})$}}
    \end{call}
    \end{sdblock}
    \begin{sdblock}{Attested Location Update}{}
    \begin{call}
    {CD}{\shortstack{\encircle{5}$S_{T}, E_{L}$}}{TS}{\encircle{6} Ack}
    \end{call}
    \end{sdblock}
    \end{sequencediagram}
    \vspace{-0.3cm}
    \caption{Secure Location Update of~\systemname{}}
    \vspace{-0.2cm}
    \label{fig:seclocupdate}
    \end{figure}

In~\systemname{}, we aim to guarantee the~\textbf{Anonymous and proximity-aware location update ($CD-S_{1}$)} property,~\ie{}, a~\ac{CD} should be able to anonymously update the location of a~\ac{TD} if and only if it is in~\ac{BLE} proximity of the~\ac{TD}. 
\fig{fig:seclocupdate} shows the sequence diagram of the protocol used to update the location of a~\ac{TD} in~\systemname{}.

First, the~\ac{CD} sends a~\texttt{Location update request} to the~\ac{TS}.
This request contains the temporary and attested public-key ($TPk_{CD}$) of the~\ac{CD} and the ID of the~\ac{TD} whose location needs to be updated.
The public-key $TPk_{CD}$ comes with a certificate chain. 
The root certificate within this chain is signed using an attestation root key,
which the device manufacturer injects into the device's hardware-backed keystore at the factory~\cite{androidsecure}. 
The~\ac{TS} responds back with two nonces, $N_{t}$ ($E_{T}$) and $N_{c}$ ($E_{C}$), which are encrypted with the public key of~\ac{TD} and~\ac{CD} respectively.

Next, as shown in the~\texttt{Sign Token} block, the~\ac{CD}, interacting with the~\ac{TEE}, decrypts $E_{C}$ to get $N_{c}$.
The nonce $N_{c}$ along with the location information (\texttt{loc}) is signed by the~\ac{TEE} with the temporary private key ($TPr_{CD}$), resulting in $L$.
The~\ac{CD} forwards the encrypted nonce $E_{T}$, $TPk_{CD}$, and $L$ to the~\ac{TD} (\encircle{3}). 
The~\ac{TD} verifies $L$ using $TPk_{CD}$ and decrypts $E_{T}$ to get the nonce $N_{t}$.
Next, the~\ac{TD} encrypts the location~\texttt{loc} and a random nonce $N_{l}$ using the location key to get $E_{L}$, and then signs both nonces, $N_{t}$ (decrypted from $E_{T}$) and $N_{c}$ (retrieved from $L$), to generate $S_{T}$.
Both $E_{L}$ and $S_{T}$ are then sent to the~\ac{CD} (\encircle{4}).
The random nonce, $N_{l}$, acts as a freshness token and avoid inference attacks. 



Finally, as shown in~\texttt{Encrypted Location Update} block, the~\ac{CD} forwards $E_{L}$ and $S_{T}$ to~\ac{TS} (\encircle{5}). The~\ac{TS} verifies $S_{T}$ to check that the signatures are valid and contain expected nonces.
Consequently, it updates its database with the location of the~\ac{TD} to be $E_{L}$.

Note that the requirement of a valid $S_{T}$ satisfies our~\capfive{}~\ac{NSC}.
In fact, to generate a valid $S_{T}$, the~\ac{CD} must communicate with the~\ac{TD}, and, hence, it has to be in~\ac{BLE} proximity.
The described location update protocol satisfies our~\textbf{Anonymous and proximity-aware location update ($CD-S_{1}$)} property.
First, the use of temporary keys ensures the anonymity of the~\ac{CD}.
Second, to perform a location update, as shown in~\texttt{Attested Location Update} block of~\fig{fig:seclocupdate}, the~\ac{CD} is expected to send $S_{T}$, which contains the nonces $N_{t} || N_{c}$ signed with the private key of the~\ac{TD}.
Thus, the~\ac{CD} can produce a valid $S_{T}$ only by communicating with the~\ac{TD}.

Furthermore, the location information of a~\ac{TD} is stored on the~\ac{TS} in encrypted form (\ie, $E_{L}$), and the decryption key is \emph{unknown to the~\ac{TS}}.
This design guarantees end-to-end encryption, protecting the private information (\ie, location) of the~\ac{TD}.

\subsection{Location query}
\label{subsec:locquery}
We want only the owners of a~\ac{TD} to be able to successfully retrieve its location from the~\ac{TS}, as this is a privileged operation.

In \systemname{}, location query by a~\ac{CD} (or user) happens by first authenticating the~\ac{CD} with the~\ac{TS}.
We use a simple challenge-response for authentication.
On successful authentication, the~\ac{TS} then checks whether the authenticated~\ac{CD} is the owner of the~\ac{TD}.
If yes, the~\ac{TS} responds back with the encrypted location token, \ie, $E_{T} = Enc(Pk_{CD}, E_{L}||N)$, where $E_{L}$ is the last-known encrypted location (\sect{subsub:locupdate}) of the requested~\ac{TD}, and, $N$ is a randomly generated nonce that acts as salt to prevent against dictionary attacks to determine whether the location has changed or not.

Then, the~\ac{CD} decrypts $E_{T}$ with its private key to obtain $E_{L}$.
The~\ac{CD} can now use the location key ($L_{TD}$), previously shared with the~\ac{TD}, to decrypt $E_{L}$ and obtain the location of the~\ac{TD}.
As mentioned in~\sect{subsubsec:primaryownercmds}, the location key is never shared with the~\ac{TS}.
Instead, such a key has to be independently shared by the primary owner of the~\ac{TD} with the other secondary owners.

In summary, our location query protocol satisfies our \textbf{Privileged owners ($TD-S_{1}$)} property, because, to perform a location query, the~\ac{CD} is expected to be the registered privileged owner of the~\ac{TD}.
Furthermore, the nonce $N$ is randomly generated for each request, thus making it impossible for a non-owner~\ac{CD} to infer whether the location of a~\ac{TD} has changed or not.


%

    
\section{Evaluation}
We implemented a preliminary prototype of~\systemname using a Raspberry Pi 3B~\cite{raspbpi3} with~\ac{BLE} as our~\ac{TD}.
We implemented the~\ac{CD} and~\ac{TS} functionalities in Python.

In this section, we evaluate the performance overhead of~\systemname in terms of execution time and energy consumption.
We measure the time overhead that affects the~\ac{CD}, as it initiates all the operations.
Whereas for energy consumption, we look at the~\ac{TD} as it has strict power requirements. Specifically, we measure how much additional energy is consumed by~\systemname compared to the baseline system.
We use a power meter~\cite{usbcpower} to measure the energy consumed by the~\ac{TD} (\ie, Raspberry Pi 3B).
We consider the unmodified TrackR system to be our baseline.

\mypar{Cryptographic operations overhead.}
We measure the overhead of the individual cryptographic operations on the~\ac{CD}, \ac{TD}, and \ac{TS}.
\tbl{tab:microtimings} shows the time required for each of the cryptographic operations along with the processors' speed.
These times are very small compared to the frequency of location update and query operations. 
In fact, almost all the~\acp{TApp} limit the frequency of such operations to once in 5 minutes.


\begin{table}[b]
\centering
\scriptsize
\caption{Execution time of the cryptographic operations (64 bytes of data with 1024-bit keys) performed on each of the entities in our implementation.}
\vspace{-0.3cm}
\label{tab:microtimings}
\begin{tabularx}{\columnwidth}{X|c c c c}
\toprule
& \multicolumn{4}{c}{\textbf{Time (ms)}}  \\
\textbf{Entity} & 
\makecell{
    \textbf{Asymmetric} \\ \textbf{Encryption}
}
&
\makecell{
    \textbf{Asymmetric} \\ \textbf{Decryption}
}
&
\makecell{
    \textbf{Symmetric} \\ \textbf{Encryption}
} &
\makecell{
    \textbf{Symmetric} \\ \textbf{Decryption}
} \\
\midrule
\textbf{TS/CD (2.6 Ghz)} & 3.0007e-03 & 2.9787e-03 & 0.23 & 0.55 \\

\textbf{TD (1.2 Ghz)} & 6.0101e-02 & 5.1594e-02 & 4.36 & 11.98 \\ 
\bottomrule
\end{tabularx}
\end{table}

\mypar{End-to-End overhead.}
We measure the end-to-end overhead introduced to our system, compared to our baseline (i.e., TrackR).
As all the operations start and end at~\ac{CD}, the timings are measured at the~\ac{CD}.
\tbl{tab:timerecoding} shows the time required by each of the operations in~\systemname compared to the baseline. 
Although the time is significantly higher compared to the baseline, the absolute numbers are still small and fall well within operational range (i.e., about 5 min).
Furthermore, additional network communication (both~\ac{BLE} and Ethernet) in~\systemname{} is the main contributor for the time overhead as the time for cryptographic operations is significantly low (\tbl{tab:microtimings}).
The power consumption of the~\ac{TD} for each operation is shown in~\tbl{tab:energyconsumed}.
Here, we multiply the average wattage of the~\ac{TD} by the total amount of time of the corresponding operation.
Thus, this represents an upper bound.
In practice, given that most of the time is spent in the network communication between the~\ac{CD} and~\ac{TS}, the actual power consumed by the~\ac{TD} can be significantly lower.
Furthermore, these numbers are again negligible if compared to the capacity of the modern batteries employed by the~\acp{TD}.


\mypar{Discussion.}
The performance numbers reported in our evaluation serve as an upper bound.
This is because, first, our implementation was not optimized and done mainly as a proof of work to test the protocols.
Second, using a Raspberry Pi 3B as a ~\ac{TD} implied that our reported energy consumption is significantly higher than an implementation on a standard \ac{BLE} device.
Finally, the adoption of specialized low power cryptographic processors~\cite{nxpcryptocoop} on the~\ac{TD} would significantly reduce the energy consumption.

%

\begin{table}[t]
\centering
\scriptsize
\caption{Average (10 runs) execution time of each operation.}
\vspace{-0.3cm}
\label{tab:timerecoding}
\begin{tabularx}{0.9\columnwidth}{X|c c c c }
\toprule
& \multicolumn{4}{c}{\textbf{Time (seconds)}} \\ 
\textbf{System}
& 
\makecell{\textbf{Owner} \\ \textbf{Registration}}
&
\makecell{\textbf{Primary Owner}\\ \textbf{Operation}}
&
\makecell{\textbf{Location} \\ \textbf{Update}}
&
\makecell{\textbf{Location} \\ \textbf{Query}} \\
\midrule
Baseline & 0.0346 & N/A & 0.0386 & 0.0002 \\
\systemname & 4.4298 & 4.7319 & 8.6322 & 0.1851 \\
\bottomrule
\end{tabularx}
\vspace{-0.2cm}
\end{table}

\begin{table}[b]
\centering
\scriptsize
\vspace{-0.2cm}
\caption{Average (10 runs) energy consumed by the~\ac{TD}.}
\vspace{-0.4cm}
\label{tab:energyconsumed}
\begin{tabularx}{0.9\columnwidth}{X|c c c c}
\toprule
& \multicolumn{4}{c}{\textbf{Energy consumed by the~\ac{TD} (Joules)}} \\ 
\textbf{System}
& 
\makecell{\textbf{Owner} \\ \textbf{Registration}}
&
\makecell{\textbf{Primary Owner}\\ \textbf{Operation}}
&
\makecell{\textbf{Location} \\ \textbf{Update}}
&
\makecell{\textbf{Location} \\ \textbf{Query}} \\
\midrule
Baseline          &  0.0005    &   N/A    &   1.66E-04  &     0     \\
\systemname       &  5.5976    &   6.7455      &   11.8824   &     0     \\
\bottomrule
\end{tabularx}
\end{table}

\section{Limitations}
\systemname{} relies on various communication protocols, such as~\ac{TLS} and~\ac{BLE} and implicitly assumes them to be secure.
Consequently,~\systemname is susceptible to any vulnerabilities~\cite{antonioli2020bias, tews2009practical, caneill2010attacks} in these protocols.
There are also other limitations of \systemname that are listed below:

\mypar{Built in capabilities.} \systemname there requires the~\acp{CD} and~\acp{TD} to have a few in-built capabilities (\sect{subsec:builtincapabilities}).
One of the important capabilities expected from a~\ac{TD} is the ability to perform cryptographic operations.
Given that~\acp{TD} are battery-powered, the cryptographic operations need to be optimized to reduce power consumption.
Low-power cryptographic processors are well-studied~\cite{cryptoforultralow, 10.1007, singh2012performance, yang2017hardware} and are used in various battery-powered devices~\cite{zhukov2015lightweight}.
These cryptographic co-processors~\cite{nxpcryptocoop} can be used to enable~\acp{TD} to perform cryptographic operations without significantly affecting their battery life.


\mypar{Relay attacks.} Similar to contact-less systems, we assume that the ability to communicate with a~\ac{TD} implies proximity.
However, attackers can leverage relay attacks~\cite{francillon2011relay} to break this assumption.
In practice, a~\ac{CD} (proxy) could serve as a relay to a~\ac{TD} and could enable other~\acp{CD} to communicate with the~\ac{TD} without being in~\ac{BLE} proximity.
Although techniques such as distance bounding~\cite{drimer2007keep}, secure distance measurement~\cite{leu2019message}, or an out-of-band hardware token~\cite{dhar2020proximitee} could be used to prevent these attacks, the low energy requirement of the~\acp{TD} makes the existing solutions impractical.

\mypar{Denial-of-Service attacks.} As a~\ac{CD} can induce a~\ac{TD} to perform cryptographic operations (\eg{}, \encircle{3},~\encircle{4} in~\fig{fig:seclocupdate}), a malicious and in-proximity~\ac{CD} could~\emph{drain the battery} of a~\ac{TD} by issuing a high number of requests.
However, these attacks can be easily prevented by implementing well-known rate-limiting techniques~\cite{wong2005effectiveness}.

\section{Related work}
The problem of effective geolocation has been well-studied.
Initial works expect the devices to have the ability to communicate with a remote server~\cite{ristenpart2008privacy} or the availability of trusted landmarks~\cite{gueye2006constraint, hu2012towards}. Both of which are not feasible in our scenario.

\mypar{Crowdsourced tracking.}
Decentralized techniques such as crowdsourcing are used to track lost objects using location-enabled smartphones~\cite{frank2007objects}.
SecureFind~\cite{sun2015securefind} provides privacy for the objects being tracked, wherein it prevents the service from accurately knowing lost devices by making the nodes (i.e., smartphones) send dummy updates to object queries.
However, in SecureFind any user can determine whether a tracker device is lost, which is a privacy concern.
Anonysense~\cite{cornelius2008anonysense} uses differential privacy techniques such as group signatures~\cite{ateniese1999some} to provide anonymity for the smartphones that upload location reports.
These techniques have high deployment overhead as they need an end-to-end infrastructure change, including trusted intermediate services.
Techu~\cite{agadakos2017techu} takes a first step toward exploring the security issues in crowdsourced location tracking.
However, the focus in Techu is mainly on the privacy of the~\acp{CD} and~\acp{TD}.
Moreover, their threat model less strict than ours and does not consider the issues of tracking device proximity and location spoofing, which could lead to Sybil attacks~\cite{wang2016defending}.
Furthermore, their solution proposes a pull model for location query, where a~\ac{CD} has to communicate with the other~\ac{CD} that performed the location update to retrieve the location.
This introduces additional attack vectors to a~\ac{CD}, as it has to accept incoming connections to perform a location update.
Recently Apple proposed FindMy service~\cite{applefindmy}, which requires the~\ac{TD} to set up an initial key pair (ECP-224) with iCloud server.
To achieve this, the TD needs to have WIFI connectivity to communicate with the iCloud server, 
and it requires an input mechanism for a user to enter her credentials.
Unfortunately, the BLE tracking devices that we study in our paper have none
of the capabilities (internet connectivity and input mechanism).
Similarly, several privacy-preserving contact tracing~\cite{troncoso2020decentralized,chan2020pact} techniques also expect geolocation and networking capabilities, which are unfortunately not available on BLE tracking devices.
Another recent work, PrivateFind~\cite{weller2020lost} performs a more detailed analysis of the security issues in Bluetooth finders.
They did an excellent job in analyzing various trackers and further, propose a privacy-friendly tracking system.
In this work, we broaden the analysis by considering~\emph{all the security aspects} of a crowdsourced tracking system.
We identify all the necessary properties (\sect{subsub:securityprop}) of a secure crowdsourced tracking system, which enable a systematic security evaluation of existing crowdsourced tracking systems (\tbl{tab:verifiedsecurityproperties}).
Furthermore, our system,~\systemname, has additional security guarantees such as resilience to Sybil attacks, which are missing in PrivateFind.

\mypar{Bluetooth attacks.}
There is a considerable amount of work that studies the security of the Bluetooth protocol~\cite{rijah2016bluetooth, jakobsson2001security}.
Recently, Antonioli et al.~\cite{antonioli2020bias} identified a series of security flaws in the specification of Bluetooth authentication and secure connection establishment, that allows attackers to impersonate Bluetooth devices.


\mypar{Geolocation and privacy.}
It is well understood that the sharing of location by a smartphone to any tracking service could pose privacy concerns~\cite{ghinita2013privacy, krumm2009survey}.
Some techniques try to protect privacy by adding noise~\cite{bordenabe2014optimal, fawaz2014location} or use differential privacy~\cite{andres2013geo, papadopoulos2010nearest, to2014framework} techniques to hinder inferring accurate location information.
We cannot use these techniques for crowdsourced tracking of BLE devices because the accuracy of the location is of real importance and is one of the primary goals of the service~\cite{friesen2015bluetooth}.
We also actively hinder the ability to spoof location information by using TrustZone~\cite{pinto2019demystifying}.
Furthermore, work from Iasonas et al.~\cite{polakis2015s} showed that it is possible to infer accurate location information even from noisy data.


\section{Conclusions}
In this work, 
we first defined and formalized a set of security properties for a generic crowdsourced location tracking system.
We proposed a new design for a secure crowdsourced tracking system,~\systemname.
Our design successfully prevents unauthorized attackers from taking advantage of the system to obtain sensitive information or arbitrarily control tracking devices.
In conclusion, we believe that \systemname will enable BLE tracking devices to utilize the benefits of the crowdsourced model while guaranteeing security and privacy.

\section*{Acknowledgements}

This material is based upon work supported by the Department of
Homeland Security (DHS) (Award No. FA8750-19-2-0005) and by the Office of Naval Research (ONR) under award number N00014-17-1-2011 and N00014-20-1-2632. Any opinions, findings, and conclusions or recommendations expressed in this publication are those of the author(s) and do not necessarily reflect the views of the ONR or DHS. 
Part of this work is supported by the Dutch Research Council (NWO) in the context of the INTERSECT research program (NWA.1160.18.301).

\bibliographystyle{ACM-Reference-Format}
\bibliography{bibs/references}


\begin{thebibliography}{70}


\ifx \showCODEN    \undefined \def \showCODEN     #1{\unskip}     \fi
\ifx \showDOI      \undefined \def \showDOI       #1{#1}\fi
\ifx \showISBNx    \undefined \def \showISBNx     #1{\unskip}     \fi
\ifx \showISBNxiii \undefined \def \showISBNxiii  #1{\unskip}     \fi
\ifx \showISSN     \undefined \def \showISSN      #1{\unskip}     \fi
\ifx \showLCCN     \undefined \def \showLCCN      #1{\unskip}     \fi
\ifx \shownote     \undefined \def \shownote      #1{#1}          \fi
\ifx \showarticletitle \undefined \def \showarticletitle #1{#1}   \fi
\ifx \showURL      \undefined \def \showURL       {\relax}        \fi
\providecommand\bibfield[2]{#2}
\providecommand\bibinfo[2]{#2}
\providecommand\natexlab[1]{#1}
\providecommand\showeprint[2][]{arXiv:#2}

\bibitem[\protect\citeauthoryear{Agadakos, Polakis, and Portokalidis}{Agadakos
  et~al\mbox{.}}{2017}]%
        {agadakos2017techu}
\bibfield{author}{\bibinfo{person}{Ioannis Agadakos}, \bibinfo{person}{Jason
  Polakis}, {and} \bibinfo{person}{Georgios Portokalidis}.}
  \bibinfo{year}{2017}\natexlab{}.
\newblock \showarticletitle{Techu: Open and privacy-preserving crowdsourced GPS
  for the masses}. In \bibinfo{booktitle}{\emph{Proceedings of the 2017 Annual
  International Conference on Mobile Systems, Applications, and Services}}.
  \bibinfo{pages}{475--487}.
\newblock


\bibitem[\protect\citeauthoryear{Amazon}{Amazon}{2020}]%
        {usbcpower}
\bibfield{author}{\bibinfo{person}{Amazon}.} \bibinfo{year}{2020}\natexlab{}.
\newblock \bibinfo{title}{USB C Power meter}.
\newblock
\newblock
\newblock
\shownote{\url{https://www.amazon.com/Tester-Eversame-Voltmeter-Ammeter-Braided/dp/B07MGQZHGM}.}


\bibitem[\protect\citeauthoryear{Andr{\'e}s, Bordenabe, Chatzikokolakis, and
  Palamidessi}{Andr{\'e}s et~al\mbox{.}}{2013}]%
        {andres2013geo}
\bibfield{author}{\bibinfo{person}{Miguel~E Andr{\'e}s},
  \bibinfo{person}{Nicol{\'a}s~E Bordenabe}, \bibinfo{person}{Konstantinos
  Chatzikokolakis}, {and} \bibinfo{person}{Catuscia Palamidessi}.}
  \bibinfo{year}{2013}\natexlab{}.
\newblock \showarticletitle{Geo-indistinguishability: Differential privacy for
  location-based systems}. In \bibinfo{booktitle}{\emph{Proceedings of the 2013
  ACM SIGSAC conference on Computer \& communications security}}.
  \bibinfo{pages}{901--914}.
\newblock


\bibitem[\protect\citeauthoryear{Antonioli, Tippenhauer, and
  Rasmussen}{Antonioli et~al\mbox{.}}{2020}]%
        {antonioli2020bias}
\bibfield{author}{\bibinfo{person}{Daniele Antonioli},
  \bibinfo{person}{Nils~Ole Tippenhauer}, {and} \bibinfo{person}{Kasper
  Rasmussen}.} \bibinfo{year}{2020}\natexlab{}.
\newblock \showarticletitle{BIAS: Bluetooth Impersonation AttackS}. In
  \bibinfo{booktitle}{\emph{Proceedings of the 2020 IEEE Symposium on Security
  and Privacy (S\&P)}}.
\newblock


\bibitem[\protect\citeauthoryear{Apple}{Apple}{2020a}]%
        {applefindmy}
\bibfield{author}{\bibinfo{person}{Apple}.} \bibinfo{year}{2020}\natexlab{a}.
\newblock \bibinfo{title}{Apple Platform Security (Page 93 Find My)}.
\newblock
\newblock
\newblock
\shownote{\url{https://manuals.info.apple.com/MANUALS/1000/MA1902/en_US/apple-platform-security-guide.pdf}.}


\bibitem[\protect\citeauthoryear{Apple}{Apple}{2020b}]%
        {applekeystore}
\bibfield{author}{\bibinfo{person}{Apple}.} \bibinfo{year}{2020}\natexlab{b}.
\newblock \bibinfo{title}{Storing Keys in the Secure Enclave}.
\newblock
\newblock
\newblock
\shownote{\url{https://developer.apple.com/documentation/security/certificate_key_and_trust_services/keys/storing_keys_in_the_secure_enclave}.}


\bibitem[\protect\citeauthoryear{Ateniese and Tsudik}{Ateniese and
  Tsudik}{1999}]%
        {ateniese1999some}
\bibfield{author}{\bibinfo{person}{Giuseppe Ateniese} {and}
  \bibinfo{person}{Gene Tsudik}.} \bibinfo{year}{1999}\natexlab{}.
\newblock \showarticletitle{Some open issues and new directions in group
  signatures}. In \bibinfo{booktitle}{\emph{Proceedings of the 1999
  International Conference on Financial Cryptography}}. Springer,
  \bibinfo{pages}{196--211}.
\newblock


\bibitem[\protect\citeauthoryear{Bordenabe, Chatzikokolakis, and
  Palamidessi}{Bordenabe et~al\mbox{.}}{2014}]%
        {bordenabe2014optimal}
\bibfield{author}{\bibinfo{person}{Nicol{\'a}s~E Bordenabe},
  \bibinfo{person}{Konstantinos Chatzikokolakis}, {and}
  \bibinfo{person}{Catuscia Palamidessi}.} \bibinfo{year}{2014}\natexlab{}.
\newblock \showarticletitle{Optimal geo-indistinguishable mechanisms for
  location privacy}. In \bibinfo{booktitle}{\emph{Proceedings of the 2014 ACM
  SIGSAC conference on computer and communications security}}.
  \bibinfo{pages}{251--262}.
\newblock


\bibitem[\protect\citeauthoryear{Caneill and Gilis}{Caneill and Gilis}{2010}]%
        {caneill2010attacks}
\bibfield{author}{\bibinfo{person}{Matthieu Caneill} {and}
  \bibinfo{person}{Jean-Loup Gilis}.} \bibinfo{year}{2010}\natexlab{}.
\newblock \showarticletitle{Attacks against the WiFi protocols WEP and WPA}.
\newblock \bibinfo{journal}{\emph{Journal, no. December}}
  (\bibinfo{year}{2010}).
\newblock


\bibitem[\protect\citeauthoryear{Cerdeira, Santos, Fonseca, and Pinto}{Cerdeira
  et~al\mbox{.}}{2020}]%
        {cerdeira2020sok}
\bibfield{author}{\bibinfo{person}{David Cerdeira}, \bibinfo{person}{Nuno
  Santos}, \bibinfo{person}{Pedro Fonseca}, {and} \bibinfo{person}{Sandro
  Pinto}.} \bibinfo{year}{2020}\natexlab{}.
\newblock \showarticletitle{SoK: Understanding the Prevailing Security
  Vulnerabilities in TrustZone-assisted TEE Systems}. In
  \bibinfo{booktitle}{\emph{Proceedings of the 2020 IEEE Symposium on Security
  and Privacy (S\&P), San Francisco, CA, USA}}. \bibinfo{pages}{18--20}.
\newblock


\bibitem[\protect\citeauthoryear{Chan, Gollakota, Horvitz, Jaeger, Kakade,
  Kohno, Langford, Larson, Singanamalla, Sunshine, et~al\mbox{.}}{Chan
  et~al\mbox{.}}{2020}]%
        {chan2020pact}
\bibfield{author}{\bibinfo{person}{Justin Chan}, \bibinfo{person}{Shyam
  Gollakota}, \bibinfo{person}{Eric Horvitz}, \bibinfo{person}{Joseph Jaeger},
  \bibinfo{person}{Sham Kakade}, \bibinfo{person}{Tadayoshi Kohno},
  \bibinfo{person}{John Langford}, \bibinfo{person}{Jonathan Larson},
  \bibinfo{person}{Sudheesh Singanamalla}, \bibinfo{person}{Jacob Sunshine},
  {et~al\mbox{.}}} \bibinfo{year}{2020}\natexlab{}.
\newblock \showarticletitle{Pact: Privacy sensitive protocols and mechanisms
  for mobile contact tracing}.
\newblock \bibinfo{journal}{\emph{arXiv preprint arXiv:2004.03544}}
  (\bibinfo{year}{2020}).
\newblock


\bibitem[\protect\citeauthoryear{Chen, Diao, Zhao, Zuo, Lin, Wang, Lau, Sun,
  Yang, and Zhang}{Chen et~al\mbox{.}}{2018}]%
        {chen2018iotfuzzer}
\bibfield{author}{\bibinfo{person}{Jiongyi Chen}, \bibinfo{person}{Wenrui
  Diao}, \bibinfo{person}{Qingchuan Zhao}, \bibinfo{person}{Chaoshun Zuo},
  \bibinfo{person}{Zhiqiang Lin}, \bibinfo{person}{XiaoFeng Wang},
  \bibinfo{person}{Wing~Cheong Lau}, \bibinfo{person}{Menghan Sun},
  \bibinfo{person}{Ronghai Yang}, {and} \bibinfo{person}{Kehuan Zhang}.}
  \bibinfo{year}{2018}\natexlab{}.
\newblock \showarticletitle{IoTFuzzer: Discovering Memory Corruptions in IoT
  Through App-based Fuzzing.}. In \bibinfo{booktitle}{\emph{Proceedings of the
  2018 NDSS}}.
\newblock


\bibitem[\protect\citeauthoryear{Cheval, Delaune, and Ryan}{Cheval
  et~al\mbox{.}}{2014}]%
        {cheval2014tests}
\bibfield{author}{\bibinfo{person}{Vincent Cheval},
  \bibinfo{person}{St{\'e}phanie Delaune}, {and} \bibinfo{person}{Mark Ryan}.}
  \bibinfo{year}{2014}\natexlab{}.
\newblock \showarticletitle{Tests for establishing security properties}. In
  \bibinfo{booktitle}{\emph{Proceedings of the 2014 International Symposium on
  Trustworthy Global Computing}}. Springer, \bibinfo{pages}{82--96}.
\newblock


\bibitem[\protect\citeauthoryear{Chipolo}{Chipolo}{2020a}]%
        {chipoloapp}
\bibfield{author}{\bibinfo{person}{Chipolo}.} \bibinfo{year}{2020}\natexlab{a}.
\newblock \bibinfo{title}{Chipolo App}.
\newblock
\newblock
\newblock
\shownote{\url{https://play.google.com/store/apps/details?id=chipolo.net.v3&hl=en_US}.}


\bibitem[\protect\citeauthoryear{Chipolo}{Chipolo}{2020b}]%
        {chipolotr}
\bibfield{author}{\bibinfo{person}{Chipolo}.} \bibinfo{year}{2020}\natexlab{b}.
\newblock \bibinfo{title}{Chipolo Tracking}.
\newblock
\newblock
\newblock
\shownote{\url{https://chipolo.net/en/}.}


\bibitem[\protect\citeauthoryear{Continella, Polino, Pogliani, and
  Zanero}{Continella et~al\mbox{.}}{2018}]%
        {continella18:bucketsec}
\bibfield{author}{\bibinfo{person}{Andrea Continella}, \bibinfo{person}{Mario
  Polino}, \bibinfo{person}{Marcello Pogliani}, {and} \bibinfo{person}{Stefano
  Zanero}.} \bibinfo{year}{2018}\natexlab{}.
\newblock \showarticletitle{There's a Hole in that Bucket! A Large-scale
  Analysis of Misconfigured S3 Buckets}. In
  \bibinfo{booktitle}{\emph{Proceedings of the 2018 ACM Annual Computer
  Security Applications Conference (ACSAC)}}.
\newblock


\bibitem[\protect\citeauthoryear{Cornelius, Kapadia, Kotz, Peebles, Shin, and
  Triandopoulos}{Cornelius et~al\mbox{.}}{2008}]%
        {cornelius2008anonysense}
\bibfield{author}{\bibinfo{person}{Cory Cornelius}, \bibinfo{person}{Apu
  Kapadia}, \bibinfo{person}{David Kotz}, \bibinfo{person}{Dan Peebles},
  \bibinfo{person}{Minho Shin}, {and} \bibinfo{person}{Nikos Triandopoulos}.}
  \bibinfo{year}{2008}\natexlab{}.
\newblock \showarticletitle{Anonysense: privacy-aware people-centric sensing}.
  In \bibinfo{booktitle}{\emph{Proceedings of the 2008 international conference
  on Mobile systems, applications, and services}}. \bibinfo{pages}{211--224}.
\newblock


\bibitem[\protect\citeauthoryear{Cube}{Cube}{2020}]%
        {cubetr}
\bibfield{author}{\bibinfo{person}{Cube}.} \bibinfo{year}{2020}\natexlab{}.
\newblock \bibinfo{title}{Cube Tracking}.
\newblock
\newblock
\newblock
\shownote{\url{https://cubetracker.com/}.}


\bibitem[\protect\citeauthoryear{Dhar, Puddu, Kostiainen, and Capkun}{Dhar
  et~al\mbox{.}}{2020}]%
        {dhar2020proximitee}
\bibfield{author}{\bibinfo{person}{Aritra Dhar}, \bibinfo{person}{Ivan Puddu},
  \bibinfo{person}{Kari Kostiainen}, {and} \bibinfo{person}{Srdjan Capkun}.}
  \bibinfo{year}{2020}\natexlab{}.
\newblock \showarticletitle{Proximitee: Hardened sgx attestation by proximity
  verification}. In \bibinfo{booktitle}{\emph{Proceedings of the 2020 ACM
  Conference on Data and Application Security and Privacy}}.
  \bibinfo{pages}{5--16}.
\newblock


\bibitem[\protect\citeauthoryear{Dierks and Rescorla}{Dierks and
  Rescorla}{2008}]%
        {dierks2008transport}
\bibfield{author}{\bibinfo{person}{Tim Dierks} {and} \bibinfo{person}{Eric
  Rescorla}.} \bibinfo{year}{2008}\natexlab{}.
\newblock \showarticletitle{The transport layer security (TLS) protocol version
  1.2}.
\newblock  (\bibinfo{year}{2008}).
\newblock


\bibitem[\protect\citeauthoryear{Drimer, Murdoch, et~al\mbox{.}}{Drimer
  et~al\mbox{.}}{2007}]%
        {drimer2007keep}
\bibfield{author}{\bibinfo{person}{Saar Drimer}, \bibinfo{person}{Steven~J
  Murdoch}, {et~al\mbox{.}}} \bibinfo{year}{2007}\natexlab{}.
\newblock \showarticletitle{Keep Your Enemies Close: Distance Bounding Against
  Smartcard Relay Attacks.}. In \bibinfo{booktitle}{\emph{Proceedings of the
  2007 USENIX security symposium}}, Vol.~\bibinfo{volume}{312}.
\newblock


\bibitem[\protect\citeauthoryear{Erickson, Paone, Paulsen, Sheets, and
  Uhlmann}{Erickson et~al\mbox{.}}{2017}]%
        {erickson2017implementing}
\bibfield{author}{\bibinfo{person}{Karl~R Erickson}, \bibinfo{person}{Phil~C
  Paone}, \bibinfo{person}{David~P Paulsen}, \bibinfo{person}{II~John~E
  Sheets}, {and} \bibinfo{person}{Gregory~J Uhlmann}.}
  \bibinfo{year}{2017}\natexlab{}.
\newblock \bibinfo{title}{Implementing hidden security key in eFuses}.
\newblock
\newblock
\newblock
\shownote{US Patent 9,570,193.}


\bibitem[\protect\citeauthoryear{Fawaz and Shin}{Fawaz and Shin}{2014}]%
        {fawaz2014location}
\bibfield{author}{\bibinfo{person}{Kassem Fawaz} {and} \bibinfo{person}{Kang~G
  Shin}.} \bibinfo{year}{2014}\natexlab{}.
\newblock \showarticletitle{Location privacy protection for smartphone users}.
  In \bibinfo{booktitle}{\emph{Proceedings of the 2014 ACM SIGSAC Conference on
  Computer and Communications Security}}. \bibinfo{pages}{239--250}.
\newblock


\bibitem[\protect\citeauthoryear{Finder}{Finder}{2020}]%
        {gofindertr}
\bibfield{author}{\bibinfo{person}{Go Finder}.}
  \bibinfo{year}{2020}\natexlab{}.
\newblock \bibinfo{title}{Go Finder HL}.
\newblock
\newblock
\newblock
\shownote{\url{https://www.gofinder.net/}.}


\bibitem[\protect\citeauthoryear{Francillon, Danev, and Capkun}{Francillon
  et~al\mbox{.}}{2011}]%
        {francillon2011relay}
\bibfield{author}{\bibinfo{person}{Aur{\'e}lien Francillon},
  \bibinfo{person}{Boris Danev}, {and} \bibinfo{person}{Srdjan Capkun}.}
  \bibinfo{year}{2011}\natexlab{}.
\newblock \showarticletitle{Relay attacks on passive keyless entry and start
  systems in modern cars}. In \bibinfo{booktitle}{\emph{Proceedings of the 2011
  Network and Distributed System Security Symposium (NDSS)}}.
  Eidgen{\"o}ssische Technische Hochschule Z{\"u}rich, Department of Computer
  Science.
\newblock


\bibitem[\protect\citeauthoryear{Frank, Bolliger, Roduner, and Kellerer}{Frank
  et~al\mbox{.}}{2007}]%
        {frank2007objects}
\bibfield{author}{\bibinfo{person}{Christian Frank}, \bibinfo{person}{Philipp
  Bolliger}, \bibinfo{person}{Christof Roduner}, {and}
  \bibinfo{person}{Wolfgang Kellerer}.} \bibinfo{year}{2007}\natexlab{}.
\newblock \showarticletitle{Objects calling home: Locating objects using mobile
  phones}. In \bibinfo{booktitle}{\emph{Proceedings of the 2007 International
  Conference on Pervasive Computing}}. Springer, \bibinfo{pages}{351--368}.
\newblock


\bibitem[\protect\citeauthoryear{Friesen and McLeod}{Friesen and
  McLeod}{2015}]%
        {friesen2015bluetooth}
\bibfield{author}{\bibinfo{person}{Marcia~R Friesen} {and}
  \bibinfo{person}{Robert~D McLeod}.} \bibinfo{year}{2015}\natexlab{}.
\newblock \showarticletitle{Bluetooth in intelligent transportation systems: a
  survey}.
\newblock \bibinfo{journal}{\emph{International Journal of Intelligent
  Transportation Systems Research}} \bibinfo{volume}{13}, \bibinfo{number}{3}
  (\bibinfo{year}{2015}), \bibinfo{pages}{143--153}.
\newblock


\bibitem[\protect\citeauthoryear{Ghinita}{Ghinita}{2013}]%
        {ghinita2013privacy}
\bibfield{author}{\bibinfo{person}{Gabriel Ghinita}.}
  \bibinfo{year}{2013}\natexlab{}.
\newblock \showarticletitle{Privacy for location-based services}.
\newblock \bibinfo{journal}{\emph{Synthesis Lectures on Information Security,
  Privacy, \& Trust}} \bibinfo{volume}{4}, \bibinfo{number}{1}
  (\bibinfo{year}{2013}), \bibinfo{pages}{1--85}.
\newblock


\bibitem[\protect\citeauthoryear{Google}{Google}{2020a}]%
        {androidkeystore}
\bibfield{author}{\bibinfo{person}{Google}.} \bibinfo{year}{2020}\natexlab{a}.
\newblock \bibinfo{title}{Hardware backed Keystore}.
\newblock
\newblock
\newblock
\shownote{\url{https://source.android.com/security/keystore}.}


\bibitem[\protect\citeauthoryear{Google}{Google}{2020b}]%
        {androidsecure}
\bibfield{author}{\bibinfo{person}{Google}.} \bibinfo{year}{2020}\natexlab{b}.
\newblock \bibinfo{title}{Verifying hardware-backed key pairs with Key
  Attestation}.
\newblock
\newblock
\newblock
\shownote{\url{https://developer.android.com/training/articles/security-key-attestation}.}


\bibitem[\protect\citeauthoryear{Gueye, Ziviani, Crovella, and Fdida}{Gueye
  et~al\mbox{.}}{2006}]%
        {gueye2006constraint}
\bibfield{author}{\bibinfo{person}{Bamba Gueye}, \bibinfo{person}{Artur
  Ziviani}, \bibinfo{person}{Mark Crovella}, {and} \bibinfo{person}{Serge
  Fdida}.} \bibinfo{year}{2006}\natexlab{}.
\newblock \showarticletitle{Constraint-based geolocation of internet hosts}.
\newblock \bibinfo{journal}{\emph{IEEE/ACM Transactions On Networking}}
  \bibinfo{volume}{14}, \bibinfo{number}{6} (\bibinfo{year}{2006}),
  \bibinfo{pages}{1219--1232}.
\newblock


\bibitem[\protect\citeauthoryear{Heydon and Hunn}{Heydon and Hunn}{2012}]%
        {heydon2012bluetooth}
\bibfield{author}{\bibinfo{person}{Robin Heydon} {and} \bibinfo{person}{Nick
  Hunn}.} \bibinfo{year}{2012}\natexlab{}.
\newblock \showarticletitle{Bluetooth low energy}.
\newblock \bibinfo{journal}{\emph{CSR Presentation, Bluetooth SIG https://www.
  bluetooth. org/DocMan/handlers/DownloadDoc. ashx}} (\bibinfo{year}{2012}).
\newblock


\bibitem[\protect\citeauthoryear{Hu, Heidemann, and Pradkin}{Hu
  et~al\mbox{.}}{2012}]%
        {hu2012towards}
\bibfield{author}{\bibinfo{person}{Zi Hu}, \bibinfo{person}{John Heidemann},
  {and} \bibinfo{person}{Yuri Pradkin}.} \bibinfo{year}{2012}\natexlab{}.
\newblock \showarticletitle{Towards geolocation of millions of IP addresses}.
  In \bibinfo{booktitle}{\emph{Proceedings of the 2012 Internet Measurement
  Conference}}. \bibinfo{pages}{123--130}.
\newblock


\bibitem[\protect\citeauthoryear{Hutter, Feldhofer, and Wolkerstorfer}{Hutter
  et~al\mbox{.}}{2011}]%
        {10.1007}
\bibfield{author}{\bibinfo{person}{Michael Hutter}, \bibinfo{person}{Martin
  Feldhofer}, {and} \bibinfo{person}{Johannes Wolkerstorfer}.}
  \bibinfo{year}{2011}\natexlab{}.
\newblock \showarticletitle{A Cryptographic Processor for Low-Resource Devices:
  Canning ECDSA and AES Like Sardines}. In
  \bibinfo{booktitle}{\emph{Proceedings of the 2011 Information Security Theory
  and Practice. Security and Privacy of Mobile Devices in Wireless
  Communication}}, \bibfield{editor}{\bibinfo{person}{Claudio~A. Ardagna} {and}
  \bibinfo{person}{Jianying Zhou}} (Eds.). \bibinfo{publisher}{Springer Berlin
  Heidelberg}, \bibinfo{address}{Berlin, Heidelberg},
  \bibinfo{pages}{144--159}.
\newblock
\showISBNx{978-3-642-21040-2}


\bibitem[\protect\citeauthoryear{Inc.}{Inc.}{2020a}]%
        {pebblebeeapp}
\bibfield{author}{\bibinfo{person}{Pebblebee Inc.}}
  \bibinfo{year}{2020}\natexlab{a}.
\newblock \bibinfo{title}{Pebblebee Finder}.
\newblock
\newblock
\newblock
\shownote{\url{https://play.google.com/store/apps/details?id=com.pebblebee.app.hive3&hl=en_US}.}


\bibitem[\protect\citeauthoryear{Inc.}{Inc.}{2020b}]%
        {tileapp}
\bibfield{author}{\bibinfo{person}{Tile Inc.}}
  \bibinfo{year}{2020}\natexlab{b}.
\newblock \bibinfo{title}{Google Play Store: Tile}.
\newblock
\newblock
\newblock
\shownote{\url{https://play.google.com/store/apps/details?id=com.thetileapp.tile&hl=en_US&gl=US}.}


\bibitem[\protect\citeauthoryear{Inc.}{Inc.}{2020c}]%
        {trackrapp}
\bibfield{author}{\bibinfo{person}{TrackR Inc.}}
  \bibinfo{year}{2020}\natexlab{c}.
\newblock \bibinfo{title}{Google Play Store: TrackR - Lost Item Tracker}.
\newblock
\newblock
\newblock
\shownote{\url{https://play.google.com/store/apps/details?id=com.phonehalo.itemtracker}.}


\bibitem[\protect\citeauthoryear{Jakobsson and Wetzel}{Jakobsson and
  Wetzel}{2001}]%
        {jakobsson2001security}
\bibfield{author}{\bibinfo{person}{Markus Jakobsson} {and}
  \bibinfo{person}{Susanne Wetzel}.} \bibinfo{year}{2001}\natexlab{}.
\newblock \showarticletitle{Security weaknesses in Bluetooth}. In
  \bibinfo{booktitle}{\emph{Proceedings of the 2001 Cryptographers’ Track at
  the RSA Conference}}. Springer, \bibinfo{pages}{176--191}.
\newblock


\bibitem[\protect\citeauthoryear{Kaps}{Kaps}{2006}]%
        {cryptoforultralow}
\bibfield{author}{\bibinfo{person}{Jens-Peter Kaps}.}
  \bibinfo{year}{2006}\natexlab{}.
\newblock \showarticletitle{Cryptography for Ultra-Low Power Devices}.
\newblock  (\bibinfo{date}{01} \bibinfo{year}{2006}).
\newblock


\bibitem[\protect\citeauthoryear{Krumm}{Krumm}{2009}]%
        {krumm2009survey}
\bibfield{author}{\bibinfo{person}{John Krumm}.}
  \bibinfo{year}{2009}\natexlab{}.
\newblock \showarticletitle{A survey of computational location privacy}.
\newblock \bibinfo{journal}{\emph{Personal and Ubiquitous Computing}}
  \bibinfo{volume}{13}, \bibinfo{number}{6} (\bibinfo{year}{2009}),
  \bibinfo{pages}{391--399}.
\newblock


\bibitem[\protect\citeauthoryear{Leu, Singh, Roeschlin, Paterson, and
  Capkun}{Leu et~al\mbox{.}}{2019}]%
        {leu2019message}
\bibfield{author}{\bibinfo{person}{Patrick Leu}, \bibinfo{person}{Mridula
  Singh}, \bibinfo{person}{Marc Roeschlin}, \bibinfo{person}{Kenneth~G
  Paterson}, {and} \bibinfo{person}{Srdjan Capkun}.}
  \bibinfo{year}{2019}\natexlab{}.
\newblock \showarticletitle{Message time of arrival codes: A fundamental
  primitive for secure distance measurement}.
\newblock \bibinfo{journal}{\emph{arXiv preprint arXiv:1911.11052}}
  (\bibinfo{year}{2019}).
\newblock


\bibitem[\protect\citeauthoryear{Li, Luo, Sun, Xia, Lu, Chen, Zang, and
  Guan}{Li et~al\mbox{.}}{2018}]%
        {li2018vbu}
\bibfield{author}{\bibinfo{person}{Wenhao Li}, \bibinfo{person}{Shiyu Luo},
  \bibinfo{person}{Zhichuang Sun}, \bibinfo{person}{Yubin Xia},
  \bibinfo{person}{Long Lu}, \bibinfo{person}{Haibo Chen},
  \bibinfo{person}{Binyu Zang}, {and} \bibinfo{person}{Haibing Guan}.}
  \bibinfo{year}{2018}\natexlab{}.
\newblock \showarticletitle{VBu on: Practical A estation of User-driven
  Operations in Mobile Apps}.
\newblock  (\bibinfo{year}{2018}).
\newblock


\bibitem[\protect\citeauthoryear{LLC.}{LLC.}{2020}]%
        {cubeapp}
\bibfield{author}{\bibinfo{person}{Cube~Tracker LLC.}}
  \bibinfo{year}{2020}\natexlab{}.
\newblock \bibinfo{title}{CUBE Tracker}.
\newblock
\newblock
\newblock
\shownote{\url{https://play.google.com/store/apps/details?id=com.blueskyhomesales.cube&hl=en_US}.}


\bibitem[\protect\citeauthoryear{Lohrke, Tajik, Krachenfels, Boit, and
  Seifert}{Lohrke et~al\mbox{.}}{2018}]%
        {lohrke2018key}
\bibfield{author}{\bibinfo{person}{Heiko Lohrke}, \bibinfo{person}{Shahin
  Tajik}, \bibinfo{person}{Thilo Krachenfels}, \bibinfo{person}{Christian
  Boit}, {and} \bibinfo{person}{Jean-Pierre Seifert}.}
  \bibinfo{year}{2018}\natexlab{}.
\newblock \showarticletitle{Key extraction using thermal laser stimulation}.
\newblock \bibinfo{journal}{\emph{IACR Transactions on Cryptographic Hardware
  and Embedded Systems}} (\bibinfo{year}{2018}), \bibinfo{pages}{573--595}.
\newblock


\bibitem[\protect\citeauthoryear{Masse}{Masse}{2011}]%
        {masse2011rest}
\bibfield{author}{\bibinfo{person}{Mark Masse}.}
  \bibinfo{year}{2011}\natexlab{}.
\newblock \bibinfo{booktitle}{\emph{REST API Design Rulebook: Designing
  Consistent RESTful Web Service Interfaces}}.
\newblock \bibinfo{publisher}{" O'Reilly Media, Inc."}.
\newblock


\bibitem[\protect\citeauthoryear{NXP}{NXP}{2020}]%
        {nxpcryptocoop}
\bibfield{author}{\bibinfo{person}{NXP}.} \bibinfo{year}{2020}\natexlab{}.
\newblock \bibinfo{title}{C29x: Crypto Coprocessor}.
\newblock
\newblock
\newblock
\shownote{\url{https://www.nxp.com/products/processors-and-microcontrollers/legacy-mcu-mpus/crypto-coprocessors/crypto-coprocessor:C29x}.}


\bibitem[\protect\citeauthoryear{Papadopoulos, Bakiras, and
  Papadias}{Papadopoulos et~al\mbox{.}}{2010}]%
        {papadopoulos2010nearest}
\bibfield{author}{\bibinfo{person}{Stavros Papadopoulos},
  \bibinfo{person}{Spiridon Bakiras}, {and} \bibinfo{person}{Dimitris
  Papadias}.} \bibinfo{year}{2010}\natexlab{}.
\newblock \showarticletitle{Nearest neighbor search with strong location
  privacy}.
\newblock \bibinfo{journal}{\emph{Proceedings of the VLDB Endowment}}
  \bibinfo{volume}{3}, \bibinfo{number}{1-2} (\bibinfo{year}{2010}),
  \bibinfo{pages}{619--629}.
\newblock


\bibitem[\protect\citeauthoryear{Pebblebee}{Pebblebee}{2020}]%
        {pebblebeetr}
\bibfield{author}{\bibinfo{person}{Pebblebee}.}
  \bibinfo{year}{2020}\natexlab{}.
\newblock \bibinfo{title}{Pebblebee Tracking}.
\newblock
\newblock
\newblock
\shownote{\url{https://pebblebee.com/}.}


\bibitem[\protect\citeauthoryear{Pinto and Santos}{Pinto and Santos}{2019}]%
        {pinto2019demystifying}
\bibfield{author}{\bibinfo{person}{Sandro Pinto} {and} \bibinfo{person}{Nuno
  Santos}.} \bibinfo{year}{2019}\natexlab{}.
\newblock \showarticletitle{Demystifying arm trustzone: A comprehensive
  survey}.
\newblock \bibinfo{journal}{\emph{ACM Computing Surveys (CSUR)}}
  \bibinfo{volume}{51}, \bibinfo{number}{6} (\bibinfo{year}{2019}),
  \bibinfo{pages}{1--36}.
\newblock


\bibitem[\protect\citeauthoryear{Polakis, Argyros, Petsios, Sivakorn, and
  Keromytis}{Polakis et~al\mbox{.}}{2015}]%
        {polakis2015s}
\bibfield{author}{\bibinfo{person}{Iasonas Polakis}, \bibinfo{person}{George
  Argyros}, \bibinfo{person}{Theofilos Petsios}, \bibinfo{person}{Suphannee
  Sivakorn}, {and} \bibinfo{person}{Angelos~D Keromytis}.}
  \bibinfo{year}{2015}\natexlab{}.
\newblock \showarticletitle{Where's wally? precise user discovery attacks in
  location proximity services}. In \bibinfo{booktitle}{\emph{Proceedings of the
  2015 ACM SIGSAC Conference on Computer and Communications Security}}.
  \bibinfo{pages}{817--828}.
\newblock


\bibitem[\protect\citeauthoryear{Raspberry}{Raspberry}{2020}]%
        {raspbpi3}
\bibfield{author}{\bibinfo{person}{Raspberry}.}
  \bibinfo{year}{2020}\natexlab{}.
\newblock \bibinfo{title}{Raspberry Pi 3B}.
\newblock
\newblock
\newblock
\shownote{\url{https://www.raspberrypi.org/products/raspberry-pi-3-model-b/}.}


\bibitem[\protect\citeauthoryear{Rijah, Mosharani, Amuthapriya, Mufthas,
  Hezretov, and Dhammearatchi}{Rijah et~al\mbox{.}}{2016}]%
        {rijah2016bluetooth}
\bibfield{author}{\bibinfo{person}{UL~Muhammed Rijah}, \bibinfo{person}{S
  Mosharani}, \bibinfo{person}{S Amuthapriya}, \bibinfo{person}{MMM Mufthas},
  \bibinfo{person}{Malikberdi Hezretov}, {and} \bibinfo{person}{Dhishan
  Dhammearatchi}.} \bibinfo{year}{2016}\natexlab{}.
\newblock \showarticletitle{Bluetooth security analysis and solution}.
\newblock \bibinfo{journal}{\emph{International Journal of Scientific and
  Research Publications}} \bibinfo{volume}{6}, \bibinfo{number}{4}
  (\bibinfo{year}{2016}), \bibinfo{pages}{333--338}.
\newblock


\bibitem[\protect\citeauthoryear{Ristenpart, Maganis, Krishnamurthy, and
  Kohno}{Ristenpart et~al\mbox{.}}{2008}]%
        {ristenpart2008privacy}
\bibfield{author}{\bibinfo{person}{Thomas Ristenpart}, \bibinfo{person}{Gabriel
  Maganis}, \bibinfo{person}{Arvind Krishnamurthy}, {and}
  \bibinfo{person}{Tadayoshi Kohno}.} \bibinfo{year}{2008}\natexlab{}.
\newblock \showarticletitle{Privacy-Preserving Location Tracking of Lost or
  Stolen Devices: Cryptographic Techniques and Replacing Trusted Third Parties
  with DHTs.}. In \bibinfo{booktitle}{\emph{Proceedings of the 2008 Usenix
  Security Symposium}}. \bibinfo{pages}{275--290}.
\newblock


\bibitem[\protect\citeauthoryear{Salomaa}{Salomaa}{2013}]%
        {salomaa2013public}
\bibfield{author}{\bibinfo{person}{Arto Salomaa}.}
  \bibinfo{year}{2013}\natexlab{}.
\newblock \bibinfo{booktitle}{\emph{Public-key cryptography}}.
\newblock \bibinfo{publisher}{Springer Science \& Business Media}.
\newblock


\bibitem[\protect\citeauthoryear{Singh and Kumar}{Singh and Kumar}{2012}]%
        {singh2012performance}
\bibfield{author}{\bibinfo{person}{Kirat~Pal Singh} {and}
  \bibinfo{person}{Dilip Kumar}.} \bibinfo{year}{2012}\natexlab{}.
\newblock \showarticletitle{Performance Evaluation of Low Power Encrypted MIPS
  Crypto Processor based on Cryptography Algorithms}.
\newblock  (\bibinfo{year}{2012}).
\newblock


\bibitem[\protect\citeauthoryear{Sun, Zhang, Jin, and Zhang}{Sun
  et~al\mbox{.}}{2015}]%
        {sun2015securefind}
\bibfield{author}{\bibinfo{person}{Jingchao Sun}, \bibinfo{person}{Rui Zhang},
  \bibinfo{person}{Xiaocong Jin}, {and} \bibinfo{person}{Yanchao Zhang}.}
  \bibinfo{year}{2015}\natexlab{}.
\newblock \showarticletitle{Securefind: Secure and privacy-preserving object
  finding via mobile crowdsourcing}.
\newblock \bibinfo{journal}{\emph{IEEE Transactions on Wireless
  Communications}} \bibinfo{volume}{15}, \bibinfo{number}{3}
  (\bibinfo{year}{2015}), \bibinfo{pages}{1716--1728}.
\newblock


\bibitem[\protect\citeauthoryear{Tews and Beck}{Tews and Beck}{2009}]%
        {tews2009practical}
\bibfield{author}{\bibinfo{person}{Erik Tews} {and} \bibinfo{person}{Martin
  Beck}.} \bibinfo{year}{2009}\natexlab{}.
\newblock \showarticletitle{Practical attacks against WEP and WPA}. In
  \bibinfo{booktitle}{\emph{Proceedings of the 2009 ACM conference on Wireless
  network security}}. \bibinfo{pages}{79--86}.
\newblock


\bibitem[\protect\citeauthoryear{Tile}{Tile}{2020}]%
        {tile}
\bibfield{author}{\bibinfo{person}{Tile}.} \bibinfo{year}{2020}\natexlab{}.
\newblock \bibinfo{title}{Tile Tracking}.
\newblock
\newblock
\newblock
\shownote{\url{https://www.thetileapp.com/en-us/}.}


\bibitem[\protect\citeauthoryear{Times}{Times}{2019}]%
        {capitalone_breach}
\bibfield{author}{\bibinfo{person}{The New~York Times}.}
  \bibinfo{year}{2019}\natexlab{}.
\newblock \bibinfo{title}{{Capital One Data Breach Compromises Data of Over 100
  Million}}.
\newblock
\newblock
\newblock
\shownote{\url{https://www.nytimes.com/2019/07/29/business/capital-one-data-breach-hacked.html}.}


\bibitem[\protect\citeauthoryear{To, Ghinita, and Shahabi}{To
  et~al\mbox{.}}{2014}]%
        {to2014framework}
\bibfield{author}{\bibinfo{person}{Hien To}, \bibinfo{person}{Gabriel Ghinita},
  {and} \bibinfo{person}{Cyrus Shahabi}.} \bibinfo{year}{2014}\natexlab{}.
\newblock \showarticletitle{A framework for protecting worker location privacy
  in spatial crowdsourcing}.
\newblock \bibinfo{journal}{\emph{Proceedings of the VLDB Endowment}}
  \bibinfo{volume}{7}, \bibinfo{number}{10} (\bibinfo{year}{2014}),
  \bibinfo{pages}{919--930}.
\newblock


\bibitem[\protect\citeauthoryear{Tracker}{Tracker}{2020}]%
        {nuttkr}
\bibfield{author}{\bibinfo{person}{Nut Tracker}.}
  \bibinfo{year}{2020}\natexlab{}.
\newblock \bibinfo{title}{NutSpace Tracking}.
\newblock
\newblock
\newblock
\shownote{\url{http://www.nutspace.com/}.}


\bibitem[\protect\citeauthoryear{TrackR}{TrackR}{2020}]%
        {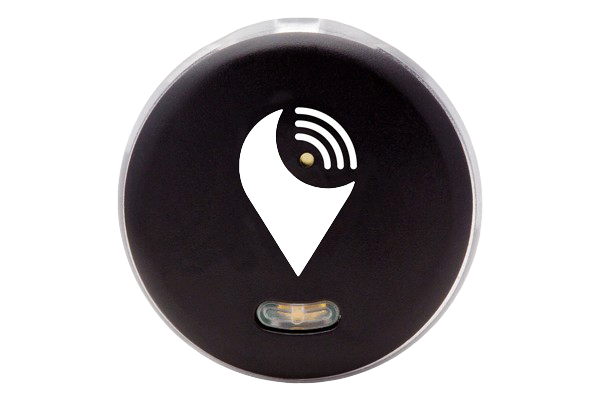}
\bibfield{author}{\bibinfo{person}{TrackR}.} \bibinfo{year}{2020}\natexlab{}.
\newblock \bibinfo{title}{TrackR Tracking}.
\newblock
\newblock
\newblock
\shownote{\url{https://www.thetrackr.com/}.}


\bibitem[\protect\citeauthoryear{Troncoso, Payer, Hubaux, Salath{\'e}, Larus,
  Bugnion, Lueks, Stadler, Pyrgelis, Antonioli, et~al\mbox{.}}{Troncoso
  et~al\mbox{.}}{2020}]%
        {troncoso2020decentralized}
\bibfield{author}{\bibinfo{person}{Carmela Troncoso}, \bibinfo{person}{Mathias
  Payer}, \bibinfo{person}{Jean-Pierre Hubaux}, \bibinfo{person}{Marcel
  Salath{\'e}}, \bibinfo{person}{James Larus}, \bibinfo{person}{Edouard
  Bugnion}, \bibinfo{person}{Wouter Lueks}, \bibinfo{person}{Theresa Stadler},
  \bibinfo{person}{Apostolos Pyrgelis}, \bibinfo{person}{Daniele Antonioli},
  {et~al\mbox{.}}} \bibinfo{year}{2020}\natexlab{}.
\newblock \showarticletitle{Decentralized privacy-preserving proximity
  tracing}.
\newblock \bibinfo{journal}{\emph{arXiv preprint arXiv:2005.12273}}
  (\bibinfo{year}{2020}).
\newblock


\bibitem[\protect\citeauthoryear{version}{version}{2020}]%
        {secrowfull}
\bibfield{author}{\bibinfo{person}{Secrow~Full version}.}
  \bibinfo{year}{2020}\natexlab{}.
\newblock \bibinfo{title}{SECROW full version}.
\newblock
\newblock
\newblock
\shownote{\url{https://drive.google.com/file/d/10tsZE_j0uCZAFd5lp6TiGZkPE6urIJPw/view?usp=sharing}.}


\bibitem[\protect\citeauthoryear{Wang, Wang, Wang, Nika, Zheng, and Zhao}{Wang
  et~al\mbox{.}}{2016}]%
        {wang2016defending}
\bibfield{author}{\bibinfo{person}{Gang Wang}, \bibinfo{person}{Bolun Wang},
  \bibinfo{person}{Tianyi Wang}, \bibinfo{person}{Ana Nika},
  \bibinfo{person}{Haitao Zheng}, {and} \bibinfo{person}{Ben~Y Zhao}.}
  \bibinfo{year}{2016}\natexlab{}.
\newblock \showarticletitle{Defending against sybil devices in crowdsourced
  mapping services}. In \bibinfo{booktitle}{\emph{Proceedings of the 2016
  Annual International Conference on Mobile Systems, Applications, and
  Services}}. \bibinfo{pages}{179--191}.
\newblock


\bibitem[\protect\citeauthoryear{Weller, Classen, Ullrich, Wa{\ss}mann, and
  Tews}{Weller et~al\mbox{.}}{2020}]%
        {weller2020lost}
\bibfield{author}{\bibinfo{person}{Mira Weller}, \bibinfo{person}{Jiska
  Classen}, \bibinfo{person}{Fabian Ullrich}, \bibinfo{person}{Denis
  Wa{\ss}mann}, {and} \bibinfo{person}{Erik Tews}.}
  \bibinfo{year}{2020}\natexlab{}.
\newblock \showarticletitle{Lost and Found: Stopping Bluetooth Finders from
  Leaking Private Information}.
\newblock \bibinfo{journal}{\emph{arXiv preprint arXiv:2005.08208}}
  (\bibinfo{year}{2020}).
\newblock


\bibitem[\protect\citeauthoryear{Wong, Bielski, Studer, and Wang}{Wong
  et~al\mbox{.}}{2005}]%
        {wong2005effectiveness}
\bibfield{author}{\bibinfo{person}{Cynthia Wong}, \bibinfo{person}{Stan
  Bielski}, \bibinfo{person}{Ahren Studer}, {and} \bibinfo{person}{Chenxi
  Wang}.} \bibinfo{year}{2005}\natexlab{}.
\newblock \showarticletitle{On the effectiveness of rate limiting mechanisms}.
  In \bibinfo{booktitle}{\emph{Proceedings of the 2005 International Symposium
  on Recent Advances in Intrusion Detection (RAID 2005)}}. Citeseer.
\newblock


\bibitem[\protect\citeauthoryear{Yang, Blaauw, and Sylvester}{Yang
  et~al\mbox{.}}{2017}]%
        {yang2017hardware}
\bibfield{author}{\bibinfo{person}{Kaiyuan Yang}, \bibinfo{person}{David
  Blaauw}, {and} \bibinfo{person}{Dennis Sylvester}.}
  \bibinfo{year}{2017}\natexlab{}.
\newblock \showarticletitle{Hardware designs for security in ultra-low-power
  IoT systems: An overview and survey}.
\newblock \bibinfo{journal}{\emph{IEEE Micro}} \bibinfo{volume}{37},
  \bibinfo{number}{6} (\bibinfo{year}{2017}), \bibinfo{pages}{72--89}.
\newblock


\bibitem[\protect\citeauthoryear{Zhukov}{Zhukov}{2015}]%
        {zhukov2015lightweight}
\bibfield{author}{\bibinfo{person}{Alexey Zhukov}.}
  \bibinfo{year}{2015}\natexlab{}.
\newblock \showarticletitle{Lightweight cryptography: modern development
  paradigms}. In \bibinfo{booktitle}{\emph{Proceedings of the 2015
  International Conference on Security of Information and Networks}}.
  \bibinfo{pages}{7--7}.
\newblock


\bibitem[\protect\citeauthoryear{Zuo, Lin, and Zhang}{Zuo
  et~al\mbox{.}}{2019}]%
        {zuo2019does}
\bibfield{author}{\bibinfo{person}{Chaoshun Zuo}, \bibinfo{person}{Zhiqiang
  Lin}, {and} \bibinfo{person}{Yinqian Zhang}.}
  \bibinfo{year}{2019}\natexlab{}.
\newblock \showarticletitle{Why does your data leak? uncovering the data
  leakage in cloud from mobile apps}. In \bibinfo{booktitle}{\emph{Proceedings
  of the 2019 IEEE Symposium on Security and Privacy (SP)}}. IEEE,
  \bibinfo{pages}{1296--1310}.
\newblock


\end{thebibliography}

\end{document}